\documentclass[11pt,a4paper]{article}
\usepackage{jheppub}
\usepackage{amsthm,bm,mathrsfs,stmaryrd}
\usepackage[table]{xcolor}
\usepackage{arydshln,tabu}
\usepackage{tikz}
\usepackage{slashed}

\newcommand{\be}{\begin{equation}}
\newcommand{\ee}{\end{equation}}
\newcommand{\ba}{\begin{eqnarray}}
\newcommand{\ea}{\end{eqnarray}}

\newcommand{\mt}[1]{$\mathop{#1}$}

\newcommand{\st}{\scriptstyle}
\newcommand{\sst}{\scriptscriptstyle}

\newcommand{\nn}{\nonumber\\}
\newcommand{\eq}{&=&}

\newcommand{\edf}{&:=&}
\newcommand{\ett}{\,\overset{\sst\rm TT}{=}\,}

\newcommand*\circled[1]{\tikz[baseline=(char.base)]{
  \node[shape=circle,draw, inner sep=1pt] (char) {#1};}}
\newcommand*\dcircled[1]{\tikz[baseline=(char.base)]{
  \node[shape=circle,fill=black!30, draw,inner sep=1pt] (char) {#1};}}

\def\sideremark#1{\ifvmode\leavevmode\fi\vadjust{\vbox to0pt{\vss% the remark
 \hbox to 0pt{\hskip\hsize\hskip1em%                          will appear only
 \vbox{\hsize3cm\tiny\raggedright\pretolerance10000%          on the side
 \noindent #1\hfill}\hss}\vbox to8pt{\vfil}\vss}}}%
\makeatletter
\newcommand{\thickhline}{%
    \noalign {\ifnum 0=`}\fi \hrule height 1pt
    \futurelet \reserved@a \@xhline
}
\newcolumntype{"}{@{\hskip\tabcolsep\vrule width 1pt\hskip\tabcolsep}}
\makeatother

\def\a{\alpha}
\def\b{\beta}
\def\g{\gamma}
\def\G{\Gamma}
\def\d{\delta}
\def\D{\Delta}
\def\e{\epsilon}

\def\l{\lambda}
\def\L{\Lambda}
\def\m{\mu}

\def\s{\sigma}
\def\S{\Sigma}
\def\t{\tau}
\def\u{\upsilon}

\def\o{\omega}
\def\O{\Omega}

\def\cD{{\cal D}}

\def\cF{{\cal F}}

\def\cO{{\cal O}}

\def\cX{{\cal X}}
\def\cY{{\cal Y}}

\author[a]{Euihun JOUNG}
\author[b]{and Massimo TARONNA}

\affiliation[a]{AstroParticule et Cosmologie\footnote{UMR 7164 
(CNRS, Universit\'e Paris 7, CEA, Observatoire de Paris)}\\ 
10 rue Alice Domon et L\'eonie Duquet, 75205 Paris Cedex 13, France}

\affiliation[b]{Max-Planck-Institut f\"ur Gravitationsphysik
(Albert-Einstein-Institut)\\
Am M\"uhlenberg 1, 14476 Golm, Germany}

\emailAdd{joung@apc.univ-paris7.fr}
\emailAdd{massimo.taronna@aei.mpg.de}

\title{\centering  
Cubic-interaction-induced deformations \\ 
of higher-spin symmetries}

\abstract{
The deformations of higher-spin symmetries induced 
by cubic interactions of symmetric massless bosonic fields
are analyzed within the metric-like formalism.
Our analysis amends the existing 
classification according to gauge-algebra deformations
taking into account also gauge-transformation deformations.
In particular, we identify a class of couplings
which leave the gauge algebra Abelian but deform one 
(out of three) gauge transformation,
and another class of couplings which deform 
all three gauge transformations in (A)dS but only two in the flat-space limit.
The former class is related to higher-spin algebra multiplets (representations of the global algebra) together with the massless-massive-massive
couplings which we also briefly discuss.
The latter class is
what makes (A)dS a distinguished background for higher-spin interactions
and 
includes in particular 
the gravitational interactions of higher-spin fields, retrospectively accounting for the Fradkin-Vasiliev solution to the Aragone-Deser problem.
We also study the restriction of gauge symmetries to global symmetries (higher-spin algebra)
discussing the invariant bilinear form and the cyclicity of the structure constants.
A possible generalization of the analysis to partially-massless fields is also 
commented.}

\begin{document}

\maketitle

\section{Introduction}

Constructing a theory of interacting higher-spin (HS) fields encounters several surprises,
such as higher derivatives, unphysical degrees of freedom or acausality --- novel obstacles which do not arise so often in their lower-spin cousins.\footnote{See \emph{e.g.} \cite{Bekaert:2010hw} (and references therein) for a review about such obstructions and their possible resolutions.}
On the other hand, there exist two examples of HS theories --- String theory and Vasiliev's equations \cite{Vasiliev:1990en,Vasiliev:1990vu,Vasiliev:1995dn,Vasiliev:2003ev,Bekaert:2005vh} --- which so far have proved to be safe from such problems.
However, the mechanisms making them bypass such difficulties, or more generally, the systematics of consistent HS interactions, are not fully understood yet.
For these reasons,
we find it important and challenging to explore this systematics in order to attain a better understanding of String theory and Vasiliev's equations, as well as to uncover possible links between them or, eventually, new theories of HS fields.

Relatively well-understood  nowadays is  the systematics of HS cubic interactions.\footnote{Many efforts have been devoted to this direction, among which 
let us mention e.g. the light-cone approach of \cite{Bengtsson:1983pd,Metsaev:1993ap,Fradkin:1995xy,Metsaev:2005ar,Metsaev:2007rn}, the metric-like approach of \cite{Berends:1984wp,Zinoviev:2008ck,Zinoviev:2010cr,Manvelyan:2010wp,
Manvelyan:2010jr,Manvelyan:2010je,
Taronna:2010qq,Sagnotti:2010at},
the frame-like approach of \cite{Fradkin:1986qy,Fradkin:1987ks,Vasilev:2011xf,Boulanger:2012dx}, the BRST-BV constructions of \cite{Boulanger:2005br,Boulanger:2008tg,Fotopoulos:2008ka,Bekaert:2010hp,Fotopoulos:2010ay,Metsaev:2012uy,Henneaux:2012wg,Henneaux:2013gba}, etc. Other relevant works are \emph{e.g.} \cite{Francia:2007qt,Francia:2008hd,Bekaert:2009ud,Bekaert:2010hk,Manvelyan:2012ww}.

Beyond the cubic order,
much less is known, in particular, 
about the (non-)local nature of the corresponding Lagrangians and most importantly about unitarity.
See \cite{Metsaev:1991mt,Metsaev:1991nb} in the light-cone formalism and \cite{Taronna:2011kt,Dempster:2012vw} 
for more recent results on quartic interactions.}
As regards the case of massless HS bosons of symmetric type, 
the exhaustive list of flat-space cubic interactions has
been identified in \cite{Bengtsson:1983pd,Metsaev:1993ap}, and classified according to the corresponding gauge-algebra deformations in \cite{Boulanger:2008tg}.
About (A)dS cubic interactions, 
all non-Abelian vertices have been constructed in \cite{Fradkin:1987ks,Fradkin:1986qy,Vasiliev:2011xf} 
in the frame-like approach (see also \cite{Boulanger:2012dx}).
In our previous works \cite{Joung:2011ww,Joung:2012rv,Joung:2012fv,Taronna:2012gb,Joung:2012hz,Joung:2013doa}, we have constructed 
all (A)dS cubic interactions in the metric-like approach,\footnote{Our 
construction covers also massless, partially-massless and massive spectrum. Note
that when considering partially-massless and massive fields, we have employed 
the gauge consistency at the level of traceless and transverse vertices, which turn
out to be equivalent to the requirement of Stueckelberg invariance. These results 
agree with those obtained in \cite{Zinoviev:2009hu}, while the study of the complete dynamics was left untouched.} 
but without touching the issues of their gauge-symmetry deformations.
In this letter, we analyze 
how the previously constructed cubic interactions 
induce deformations of the gauge transformations and gauge algebras. 
This reveals some interesting structures of HS gauge interactions.
For instance, although the total number of consistent \mt{s_{1}\!-\!s_{2}\!-\!s_{3}} couplings
is the same in flat and in (A)dS spaces,
the corresponding deformation of gauge symmetries 
is qualitatively different.
Postponing a detailed account to the next sections,
let us just comment that this is a general 
mechanism behind the Fradkin-Vasiliev construction \cite{Fradkin:1986qy,Fradkin:1987ks}
where (A)dS backgrounds play a distinguished role.

\subsection{General procedure}

For a more concrete understanding, let us remind the general program
\cite{Berends:1984rq}.
The starting point is the expansion of the sought action 
and gauge transformations in powers of the fields:
\be
	S=S^{\sst (2)}+S^{\sst (3)}+\cdots\,,\qquad
	\delta_{\varepsilon}\,\varphi=\delta^{\sst (0)}_{\varepsilon}\varphi
	+\delta^{\sst (1)}_{\varepsilon}\varphi+\cdots\,.
\ee
Here, the superscript $(n)$ means that the corresponding term involves $n$-th powers of the fields.
In this expansion scheme, the gauge invariance of the action 
is recast into an infinite number of coupled equations:
\be
	\delta_{\varepsilon}\,S=0
	\quad\Rightarrow\quad
	\left\{\begin{array}{cc}
	\delta^{\sst (0)}_{\varepsilon}\,S^{\sst (2)}=0 &\quad \circled{\scriptsize 0}\\
	\delta^{\sst (0)}_{\varepsilon}\,S^{\sst (3)}
	+\delta^{\sst (1)}_{\varepsilon}\,S^{\sst (2)}=0 & 
	\quad  \circled{\scriptsize 1}\\
	\delta^{\sst (0)}_{\varepsilon}\,S^{\sst (4)}
	+\delta^{\sst (1)}_{\varepsilon}\,S^{\sst (3)}
	+\delta^{\sst (2)}_{\varepsilon}\,S^{\sst (2)}=0 & \quad 
	\circled{\scriptsize 2}\\
	\vdots
	\end{array}
	\right.,
	\label{gauge inv}
\ee
where $S^{\sst (2)}$ and $\delta^{\sst (0)}_{\varepsilon}\varphi$ are the free action
and the corresponding gauge transformations, respectively.
Higher-order parts of the action, $S^{\sst (n\ge3)}$\,, and higher-order gauge transformations, 
$\delta^{\sst (n\ge1)}_{\varepsilon}\varphi$\,, can be identified starting from 
$S^{\sst (2)}$ and 
$\delta^{\sst (0)}_{\varepsilon}\varphi$ by solving the above equations. The strategy is to solve the equations $\circled{\tiny n}$ of \eqref{gauge inv} in two steps as
\be
	S^{\sst (2)}\,,\, \delta^{\sst (0)}_{\varepsilon}\varphi
	\ \ \overset{\dcircled{\tiny 1}}{\bm\longrightarrow} \ \ 
	S^{\sst (3)}
	\ \ \overset{\circled{\tiny 1}}{\bm\longrightarrow} \ \
	\delta^{\sst (1)}_{\varepsilon}\varphi
	\ \ \overset{\dcircled{\tiny 2}}{\bm\longrightarrow} \ \
	S^{\sst (4)}
	\ \ \overset{\circled{\tiny 2}}{\bm\longrightarrow} \ \
	\delta^{\sst (2)}_{\varepsilon}\varphi
	\ \ {\bm\longrightarrow} \ \
	\cdots\,,
\ee
where $\dcircled{\tiny n}$ represents the same condition as $\circled{\tiny n}$ but solved this time
on the shell of \emph{free} EoM. 
In particular, at each order one can first solve for 
$S^{\sst (n+2)}$ using  $\dcircled{\tiny n}$
and then read off $\delta^{\sst (n)}_{\varepsilon}\varphi$
from $\circled{\tiny n}$\,.
On the other hand, 
the full non-linear gauge transformations must form an (open) algebra: 
\be
	\delta_{\varepsilon_{1}}\,\delta_{\varepsilon_{2}}\,\varphi
	-\delta_{\varepsilon_{2}}\,\delta_{\varepsilon_{1}}\,\varphi
	= \delta_{[\![\varepsilon_{1},\varepsilon_{2}]\!]}\,\varphi
	+{\rm (trivial)}\,,
\ee
{where (trivial) refers to the trivial symmetry\footnote{A trivial symmetry of an action $S$ is given in de-Witt notation by $\delta \varphi^{i}=C^{ij}\,\delta S/\delta \varphi^{j}$
with $C^{ij}=-C^{ji}$\,, and is proportional to the EoM by definition.}
 of the action,} while the commutator $[\![\varepsilon_{1},\varepsilon_{2}]\!]$ is 
in principle field-dependent and can be expanded as
\be
	[\![\,\varepsilon_{1}\,,\,\varepsilon_{2}\,]\!]
	=[\![\,\varepsilon_{1}\,,\,\varepsilon_{2}\,]\!]^{\sst (0)}+
	[\![\,\varepsilon_{1}\,,\,\varepsilon_{2}\,]\!]^{\sst (1)}+\cdots\,.
\ee
For the purpose of the present letter, 
we focus on the lowest-order part of the commutator:
\be
	\delta^{\sst (0)}_{\varepsilon_{1}}\,
	\delta^{\sst (1)}_{\varepsilon_{2}}\,\varphi
	-\delta^{\sst (0)}_{\varepsilon_{2}}\,\delta^{\sst (1)}_{\varepsilon_{1}}\,
	\varphi
	=\delta^{\sst (0)}_{[\![\varepsilon_{1},\varepsilon_{2}]\!]^{(0)}}\varphi\,,
\ee
which is field-independent and can be entirely obtained from $\delta^{\sst (1)}_{\varepsilon}$\,.
To sum up, once consistent cubic interactions are determined for a given free theory,
then they induce deformations
of the gauge transformations and also of the gauge algebra: 
\be
	S^{\sst (3)}\quad \Rightarrow \quad \delta^{\sst (1)}_{\varepsilon}\varphi
	\quad \Rightarrow
	\quad [\![\,\varepsilon_{1}\,,\,\varepsilon_{2}\,]\!]^{\sst (0)}\,.
\ee
For the analysis of such deformations, we need first to define the free theory.
In the following, we shall consider symmetric massless bosonic  fields in
a constant curvature background, namely (A)dS.  

\subsubsection*{Ambient-space formulation and 
Transverse and Traceless part}

In order to  conveniently treat fields in (A)dS, we use the ambient-space formulation where fields 
$\varphi_{\mu_{1}\cdots\mu_{s}}(x)$ are described through the corresponding ambient avatars $\Phi_{\sst M_{1}\cdots M_{s}}(X)$, defined in a $(d+1)$-dimensional flat space and subject to homogeneity and tangentiality conditions:
\be
	(X\cdot\partial_{X}-U\cdot\partial_{U}+2+\mu)\,\Phi(X,U)=0\,,
	\qquad
	X\cdot \partial_{U}\,\Phi(X,U)=0\,.
	\label{HT}
\ee
The mass of the field is here parametrized by the degree of homogeneity $\mu$\,,
and when the field is massless (that is $\mu=0$), it admits gauge symmetries:
\be
	\delta^{\sst (0)}_{\sst E}\,\Phi
	=U\cdot \partial_{X}\,E
	\qquad
	\big[\,\partial_{U}^{2}\,E=0\,,\ 
	(X\cdot\partial_{X}-U\cdot\partial_{U})E=0\,,
	\ 
	X\cdot\partial_{U}\,E=0\,\big]\,.
	\label{g tr}
\ee
In constructing gauge-invariant interaction vertices, 
we focus for simplicity on the transverse and traceless (TT) part of the latter, disregarding the 
terms proportional to divergences and traces of the fields.
This is equivalent to consider,
instead of the full vertices and their gauge variations,
their quotient modulo the following equivalence relations:  
\be
	\partial_{U}\!\cdot \partial_{X}\,\Phi \ett 0\,,
	\qquad
	\partial_{U}^{\,2}\,\Phi \ett 0\,;
	\qquad
	\partial_{U}\!\cdot \partial_{X}\,E \ett 0\,,
	\qquad
	\partial_{X}^{\,2}\,E \ett 0\,.
	\label{TT}
\ee
In this setting, the free action assumes a general form and is given simply by
\be
	S^{\sst (2)}[\Phi]\ett -\,\frac12\
	\int_{\rm\sst (A)dS}
	e^{\partial_{U_{1}}\!\cdot\,\partial_{U_{2}}}\,
	\Phi(X,U_{1})\,\partial_{X}^{\,2}\,\Phi(X,U_{2})\,
	\Big|_{\sst U_{i}=0}\,,
	\label{amb free}
\ee
and the interaction parts of action $S^{\sst (n\ge 3)}[\Phi]$ together with 
the corresponding
gauge transformations $\delta^{\sst (n\ge1)}_{\sst E}\,\Phi$
can be also studied 
within this description. 

{Let us emphasize that here we do \emph{not} impose any 
gauge condition on the theory, but we merely attempt to
identify the TT part of interacting vertices ---
that is the reason why we introduce the notion of the equivalence
class $\ett$.
The point is, as shown in \cite{Joung:2012fv} and also recalled in the Appendix \ref{sec: TT}, that it is possible to determine such TT part
without having any information on the other parts of the vertices.}

\subsection{Summary of our results}

The interaction parts of the action can be conveniently expressed as
\be
	S^{\sst (n)}[\Phi] \ett \int_{\sst\rm (A)dS}
	C^{\sst (n)}\ \Phi(X_{1},U_{1})\,\cdots\,\Phi(X_{n},U_{n})\,\Big|_{{}^{X_{i}=X}_{U_{i}=0}}\,,
	\label{n-th order}
\ee
in terms of a differential operator $C^{\sst(n)}$ in $X_{i}$ and $U_{i}$\,.
Each operator $C^{\sst(n)}$ is constrained by the gauge-invariance conditions \eqref{gauge inv}.
For cubic interactions ($n=3$), the general solution 
to them was found to be
\be
	C^{\sst (3)}=
	\!\sum_{s_{1},s_{2},s_{3},n}\! k^{s_{1}s_{2}s_{3}}_{n}\,
	P^{\sst [n]}_{s_{1}s_{2}s_{3}}\,,\qquad
	P^{\sst [n]}_{s_{1}s_{2}s_{3}}=e^{\l\,\cD}\,Y_{1}^{\,s_{1}-n}\,Y_{2}^{\,s_{2}-n}\,Y_{3}^{\,s_{3}-n}\,G^{n}\,,
\ee
where the $s_{i}$'s are the spins of the fields involved in the interactions 
and \mt{n=0,1,\ldots, s_{\rm \sst min}} labels the possible interactions
for a given set of fields. 
The differential operator  $\cD$ acting on $Y_{i}$'s and $G$ generates 
the curvature (treated here as $\lambda$) corrections, and  its form will be provided later. 
The $Y_{i}$'s and $G$ are given by
\be
	Y_{i}=\partial_{U_{i}}\!\cdot\partial_{X_{i+1}}\,,\qquad
	G=\partial_{U_{1}}\!\cdot\partial_{U_{2}}\,\partial_{U_{3}}\!\cdot\partial_{X_{1}}
	+\partial_{U_{2}}\!\cdot\partial_{U_{3}}\,\partial_{U_{1}}\!\cdot\partial_{X_{2}}
	+\partial_{U_{3}}\!\cdot\partial_{U_{1}}\,\partial_{U_{2}}\!\cdot\partial_{X_{3}}\,,
\ee
and they represent different tensor structures allowed by gauge invariance.

Each of the cubic interactions $P^{\sst [n]}_{s_{1}s_{2}s_{3}}$
induces a deformation of gauge transformation 
$\delta^{\sst (1)}_{\sst E}\Phi$ 
and gauge algebra
$[\![\,E_{1}\,,\,E_{2}\,]\!]^{\sst (0)}$\,.
However, the main question is to analyze if those are \emph{trivial}, being a mere consequence of 
gauge-field and/or gauge-parameter redefinitions.
Hence, in order to identify the non-trivial ones, we need to analyze  
the effects of redefinitions on the deformations of gauge symmetries.\footnote{By gauge symmetries, we shall refer to both gauge transformations
and gauge algebras.}
The latter analysis is one of the main contents of the present article,
and the results are summarized in Table \ref{tab: classification}.
\begin{table}[h]
\centering
\begin{tabular}{ |c||c| c || c | c | c || c|}
  \hline                    
  &$n$ & $\#_{\partial}$ & $\delta^{\sst (1)}_{\sst E_{1}}$ & $\delta^{\sst (1)}_{\sst E_{2}}$ 
  & $\delta^{\sst (1)}_{\sst E_{3}}$ & 
  $C^{\sst (3)}$\\[2pt]
  \hline
&0 & $s_{1}+s_{2}+s_{3}$  & \cellcolor{blue!25}$=0$  & \cellcolor{blue!25}$=0$ & \cellcolor{blue!25}$=0$  & 
 \\	  
Class I &\vdots  &$\vdots$ & \cellcolor{blue!25}\vdots & \cellcolor{blue!25}\vdots & \cellcolor{blue!25}\vdots & 
$\approx \tilde K(Y_\ell, H_{12},H_{23},H_{31})$ \\ 
 & $\tfrac{s_{2}+s_{3}-s_{1}}2$  &$ 2\,s_{1}$ & \cellcolor{blue!25}$=0$ & \cellcolor{blue!25}\vdots & \cellcolor{blue!25}\vdots &
 $\ell=2$ or $3$   \\
 \hdashline
&\vdots  &  $\vdots$ & \cellcolor{red!25}$\neq0$ & \cellcolor{blue!25}\vdots & \cellcolor{blue!25}\vdots &   \\ 
Class II&\vdots  &  $\vdots$ & \cellcolor{red!25}$\vdots$ & \cellcolor{blue!25}\vdots & \cellcolor{blue!25}\vdots & $\approx 
\tilde K(Y_{1},H_{12}, H_{23}, H_{31})$ \\ 
&$\tfrac{s_{3}+s_{1}-s_{2}}2$  &   $ 2\,s_{2}$ &  \cellcolor{red!25}\vdots &  \cellcolor{blue!25}$=0$ & \cellcolor{blue!25}$=0$ &  \\
\hdashline 
&\vdots  &  $\vdots$ & \cellcolor{red!25}\vdots &  \cellcolor{red!25}$\neq 0$ &  \cellcolor{yellow!30}$\L$  & \cellcolor{black!15}  \\ 
Class III&\vdots  &  $\vdots$ & \cellcolor{red!25}\vdots &  \cellcolor{red!25}\vdots &  \cellcolor{yellow!30}$\vdots$ & \cellcolor{black!15} \\ 
&$\tfrac{s_{1}+s_{2}-s_{3}}2$  &  $2s_{3}$ & \cellcolor{red!25}\vdots & \cellcolor{red!25}\vdots & \cellcolor{yellow!30}$\L$  & 
\cellcolor{black!15} \\ 
 \hdashline 
&\vdots  &  $\vdots$ & \cellcolor{red!25}\vdots &  \cellcolor{red!25}\vdots &  \cellcolor{red!25}$\neq0$  & \cellcolor{black!15}  \\ 
Class IV&\vdots  &  $\vdots$ & \cellcolor{red!25}\vdots &  \cellcolor{red!25}\vdots &  \cellcolor{red!25}$\vdots$ & \cellcolor{black!15} \\ 
$\phantom{\Big|}$ &$s_{3}$  &  $s_{1}+s_{2}-s_{3}$ & \cellcolor{red!25}$\neq0$ & \cellcolor{red!25}$\neq0$ & \cellcolor{red!25}$\neq0$ & 
\cellcolor{black!15} \\ 
\hline    
\end{tabular}
\caption{Classification of cubic interactions ($s_{1}\ge s_{2} \ge s_{3}$) according to the deformations of gauge transformations. 
Here $\#_{\partial}$ is the number of (highest, in the (A)dS case) derivatives involved in $C^{\sst (3)}$\,.
Notice that depending on the choice of $s_{1}\!-\!s_{2}\!-\!s_{3}$, some of the above classes can be empty.}
\label{tab: classification}
\end{table}
An important lesson is that cubic interactions 
corresponding to \emph{trivial} deformations of gauge symmetries
are related to the appearance of a new tensor structure $H_{ij}$\,:
\be
	H_{ij}
	=\partial_{U_{i}}\!\cdot\partial_{X_{j}}\,
	\partial_{U_{j}}\!\cdot\partial_{X_{i}}
	-\partial_{X_{i}}\!\cdot\partial_{X_{j}}\,\partial_{U_{i}}\!\cdot\partial_{U_{j}}\,.
	\label{H}
\ee
The $H_{ij}$'s are operators taking the curls of 
the $i$-th and $j$-th fields and contracting them.
They are gauge invariant
without making use of the on-shell condition, and hence,
they do not lead to any deformation of the gauge transformations. It turns out that for \mt{s_{1}\ge s_{2}\ge s_{3}} the couplings can be organized as follows:
\begin{itemize}
\item {\bf Class I}\,:\ 
the couplings which can be re-expressed
as a function of $H_{ij}$'s
and $Y_{2}$ (or $Y_{3}$)
do not deform any of the gauge transformations $\delta^{\sst (1)}_{\sst E_{1}}$\,,
$\delta^{\sst (1)}_{\sst E_{2}}$ and $\delta^{\sst (1)}_{\sst E_{3}}$;
\item {\bf Class II}\,:\ 
the couplings which can be re-expressed as a function of $H_{ij}$'s
and $Y_{1}$ do not deform the gauge transformations $\delta^{\sst (1)}_{\sst E_{2}}$ and $\delta^{\sst (1)}_{\sst E_{3}}$ but \mt{\delta^{\sst (1)}_{\sst E_{1}}}\,;
\item {\bf Class III \& IV}\,:\ 
the couplings which cannot be re-expressed 
as a function of $H_{ij}$'s as above always deform 
all the gauge transformations in (A)dS (in flat space,
the couplings belonging to Class III deform two of the gauge transformations while those belonging to Class IV deform all of them).
\end{itemize}
As regards the gauge-algebra deformations, 
the classification can be stated as follows:
\begin{itemize}
\item {\bf Gauge algebra}\,:\ 
the deformation of the bracket $[\![ E_{1},E_{2} ]\!]^{\sst (0)}_3$ is non-trivial
if and only if both of $\delta^{\sst (1)}_{\sst E_{1}}$ and $\delta^{\sst (1)}_{\sst E_{2}}$
are non-trivial.
\end{itemize}
Let us remark that the couplings of Class II 
make the fields $\Phi_{2}$ and $\Phi_{3}$ charged with respect to $\Phi_{1}$\,,
so they are relevant for HS-algebra multiplets and in general for the classification of representations of HS algebra.
The couplings of Class III induce non-trivial deformations $\delta^{\sst (1)}_{\sst E_{3}}$\,,
but it vanishes in the flat-space limit.
This class includes in particular 
the lowest-derivative \mt{s\!-\!s\!-\!2} couplings, namely the gravitational interactions of HS fields: in (A)dS, the corresponding $\delta^{\sst (1)}_{\sst E_{3}}$ reproduces the general coordinate covariance
for HS fields, while in the flat-space limit, the latter covariance is lost, so the spin 2 field is found to not couple \emph{gravitationally} to HS fields.
Therefore, one recognizes the Aragone-Deser problem \cite{Aragone:1979hx}
and the Fradkin-Vasiliev solution \cite{Fradkin:1987ks,Fradkin:1986qy} in 
this framework.

Since any cubic interaction can be expressed in terms of
a coupling between a gauge field and a current,
it may be also useful to restate the classification above in terms of currents.
In both Class I and II cases,
the corresponding currents are gauge invariant being bilinear in the HS curvature tensors:
generalized Bell-Robinson currents.
More in details, they are
improvements or genuine Noether currents
depending on whether the corresponding couplings
belong to Class I or II.
On the other hand, the currents associated with the couplings belonging to Class III-IV are not gauge invariant.
Let us also note that the couplings belonging to Class I, II 
and III-IV correspond respectively to the \emph{Abelian}, \emph{current} 
and \emph{non-Abelian} couplings studied in \cite{Vasiliev:2011xf} within the frame-like formalism.

\subsection{Organization of paper}

The paper is organized as follows. 
In Section~\ref{sec: cubic} we review the construction of metric-like HS cubic interactions. 
In Section~\ref{sec: gaugedeform}
we analyze the non-trivial deformations of gauge-transformation deformations arriving to Table~\ref{tab: classification}. 
In Section~\ref{sec: gaugealgebra} we analyze 
the gauge-algebra deformations, 
while in Section~\ref{sec: Global} we consider their restriction to global symmetries. 
In Section~\ref{sec: GH} we give some details on the explicit structure of the non-deforming couplings in (A)dS.
Our conclusions, together with discussions on the partially-massless extension of the classification, are presented in Section~\ref{sec: Conclusions}.
The appendices contain  some technical details.

\section{Review: cubic interactions}
\label{sec: cubic}

For completeness and in order to fix the notations, we briefly summarize our previous results.
For cubic interactions, that is the \mt{n=3} case of \eqref{n-th order}, 
the ansatz\footnote{Here, we use the convention that all the fields entering in the interactions are treated as different fields, so that no symmetry under 
field-label interchange is assumed. 
This convention makes more transparent the analysis of the present work, while it is equivalent to the convention of a single generating function of HS fields. 
Hence, the results obtained in this convention can also account for the self-interaction cases.} can be further simplified, removing the ambiguities 
of integration by parts and field redefinitions, as
\be
	S^{\sst (3)}[\Phi_{1},\Phi_{2},\Phi_{3}] \ett
	\int_{\sst\rm (A)dS}
	C(Y,Z)\,\Phi_{1}(X_{1},U_{1})\ \Phi_{2}(X_{2},U_{2})\
	\Phi_{3}(X_{3},U_{3})\,\Big|_{\overset{X_{i}=X}{\sst U_{i}=0}}\,,
	\label{amb int}
\ee
where $C(Y,Z)$ is a polynomial function of six variables:
\be
	Y_{i}:=\partial_{U_{i}}\!\cdot\partial_{X_{i+1}}\,,\qquad
	Z_{i}:=\partial_{U_{i+1}}\!\!\cdot\partial_{U_{i-1}}\qquad [i\simeq i+3]\,.
\ee
When one of the fields, say $\Phi_{1}$, is massless, 
the cubic interaction must be compatible with the corresponding gauge symmetry:  \mt{\delta^{\sst (0)}_{\sst E_{1}}\,S^{\sst (3)}
	+\delta^{\sst (1)}_{\sst E_{1}}\,S^{\sst (2)}=0}\,.
This condition implies a weaker condition:
\mt{\delta^{\sst (0)}_{\sst E_{1}}\,S^{\sst (3)}\approx 0}\,,
where $\approx$ is henceforth the equivalence modulo free field equations 
as well as traces and divergences.
The latter condition can be translated into the following differential equation:
\be\label{PDE}
	\left[Y_{2}\,\partial_{Z_{3}}-Y_{3}\,\partial_{Z_{2}}-\l
	\left(Y_{2}\,\partial_{Y_{2}}-Y_{3}\,\partial_{Y_{3}}
	+\tfrac{\mu_{2}-\mu_{3}}2
	\right)\partial_{Y_{1}}\right]
	C(Y,Z)=0\,.
\ee
Here, $\l$ is an auxiliary variable introduced to simplify the computation: it eventually generates the cosmological constant $\L$
with some dimension dependent factor
 (see Appendix \ref{App: A} for more detail).

In the case of cubic interactions,
where all three fields are massless: \mt{\mu_{1}=\mu_{2}=\mu_{3}=0}\,, 
the general solution
to the cubic-interaction problem is given by
\be
	C(Y,Z)
	=e^{\l\,\cD}\,
	K(Y_{1},Y_{2},Y_{3},G)\,\Big|_{G=G(Y,Z)},
	\label{massless sol}
\ee
where 
$\cD$ and $G(Y,Z)$ are defined by
\ba
	&&\cD:=Z_{1}\partial_{Y_{2}}\partial_{Y_{3}}+\,
	Z_{1}Z_{2}\partial_{Y_{3}}\partial_{G}\,+\,{\rm cyc.}\,
	+ \,Z_{1}Z_{2}Z_{3}\partial_{G}^{\,2}\,,
	\label{D op}
	\\
	&&G(Y,Z):=Y_{1}\,Z_{1}+Y_{2}\,Z_{2}+Y_{3}\,Z_{3}\,.
\ea
From the solution \eqref{massless sol},
one can extract the $s_{1}\!-\!s_{2}\!-\!s_{3}$ interactions
by Taylor-expanding the function $K$,
and the contribution of each monomial in $K$ to the cubic interaction $C$ reads
(for $n=0,1,\ldots, s_{\rm\sst min}$)
\ba
	P^{\sst [n]}_{s_{1}s_{2}s_{3}}(Y,Z)\edf
	e^{\l\,\cD}\,
	Y_{1}^{s_{1}-n}\,Y_{2}^{s_{2}-n}\,Y_{3}^{s_{3}-n}\,G^{n}
	\,\Big|_{G=G(Y,Z)}\nn
	\eq 
	Y_{1}^{s_{1}-n}\,Y_{2}^{s_{2}-n}\,Y_{3}^{s_{3}-n}\,[G(Y,Z)]^{n}
	+\cO(\l)\,.
	\label{Cn}
\ea
The first term in the second line
corresponds to the highest-derivative part of coupling
while $\mathcal O(\l)$ represents the lower-derivative terms.
So, the highest number, $\#_{\partial}$\,, of derivatives in $P^{\sst [n]}_{s_{1}s_{2}s_{3}}$ is 
\be
	\#_{\partial}=s_{1}+s_{2}+s_{3}-2n\,,
	\label{N derive}
\ee
and the corresponding part coincides with the flat-space coupling in the 
$\L\to 0$ limit.

\section{Deformations of gauge transformations and Non-deforming couplings}
\label{sec: gaugedeform}

\subsection{Deformations of gauge transformations}

After constructing all consistent cubic interactions $S^{\sst (3)}$, 
the next step is to identify the corresponding deformations
of the gauge transformations $\delta^{\sst (1)}_{\sst E_{i}}$ from 
the gauge-invariance condition:
\be\label{1st con}
	\delta^{\sst (1)}_{\sst E_1}\left(S^{\sst (2)}[\Phi_{2}]
	+S^{\sst (2)}[\Phi_{3}]\right)
	+\delta^{\sst (0)}_{\sst E_{1}}\,S^{\sst (3)}[\Phi_{1},\Phi_{2},\Phi_{3}]=0\,.
\ee
Here, without loss of generality, we have chosen $E_{i}=E_{1}$. 
Using the form \eqref{amb free} of the quadratic action, 
the above condition can be put into
\ba
	&& \delta^{\sst (0)}_{\sst E_{1}}\,S^{\sst (3)}[\Phi_{1},\Phi_{2},\Phi_{3}] \nn
	&&\ett
	 \int_{\rm\sst (A)dS} 
	e^{\partial_{U_{1}}\!\cdot\,\partial_{U_{2}}}\Big[
	\delta^{\sst (1)}_{\sst E_{1}}\Phi_{2}(X,U_{1})\,\partial_{X}^{\,2}\,\Phi_{2}(X,U_{2})
	+\delta^{\sst (1)}_{\sst E_{1}}\Phi_{3}(X,U_{1})\,\partial_{X}^{\,2}\,\Phi_{3}(X,U_{2})\Big]_{\sst U_{i}=0}.\qquad
	\label{offshell cond}
\ea
As one can see from this expression,
the deformation $\delta^{\sst (1)}_{\sst E_{1}}$ can be obtained 
by computing the \emph{off-shell} variation of the cubic interactions
\eqref{amb int}.
The latter reads
\ba
	&&\delta^{\sst (0)}_{\sst E_{1}}\,S^{\sst (3)}[\Phi_{1},\Phi_{2},\Phi_{3}]\nn
	&&\ett \int_{\sst\rm (A)dS}
	\partial_{Y_{1}}C(Y,Z)\,
	E_{1}(X_{1},U_{1})\,\frac12\left(\partial_{X_{3}}^{2}-\partial_{X_{2}}^{2}
	\right)
	\Phi_{2}(X_{2},U_{2})\,\Phi_{3}(X_{3},U_{3})
	\,\Big|_{\overset{X_{i}=X}{\sst U_{i}=0}}\,,\qquad
	\label{f var tot}
\ea
where $C$ is the solution given in eq.~\eqref{massless sol}.
In order to obtain the deformations $\delta^{\sst (1)}_{\sst E_{1}}\Phi_{2}$
and $\delta^{\sst (1)}_{\sst E_{1}}\Phi_{3}$\,,
we need to recast eq.~\eqref{f var tot} into the form of eq.~\eqref{offshell cond}. 
{ More precisely, we massage
 the terms proportional to 
$\partial_{X_{3}}^{2}$ and $\partial_{X_{2}}^{2}$ in eq.~\eqref{f var tot}
to get the expressions for $\delta^{\sst (1)}_{\sst E_{1}}\Phi_{3}$
and $\delta^{\sst (1)}_{\sst E_{1}}\Phi_{2}$\,, respectively.
Herefrom, let us focus on $\delta^{\sst (1)}_{\sst E_{1}}\Phi_{3}$\,, because 
$\delta^{\sst (1)}_{\sst E_{1}}\Phi_{2}$ can be obtained from the former 
by interchanging the field labels. 
What one needs to do is to integrate  by parts all the $\partial_{X_{3}}$'s in $\partial_{Y_{1}}C(Y,Z)$ of 
eq.~\eqref{f var tot} because in \eqref{offshell cond} there is no derivative
acting on $\partial_{X}^{2}\,\Phi_{3}$\,.
Notice that, since the variable $Y_{2}$ involves $\partial_{X_{3}}$\,,
we can no more use such a variable in $C(Y,Z)$
and this requires to re-express $C(Y,Z)$ in terms of 
new variables $\bar Y$\,:
\be
	 \bar Y_{1}:=\partial_{U_{1}}\!\!\cdot\partial_{X_{2}}\,,\quad
	\bar Y_{2}:=-\,\partial_{U_{2}}\!\!\cdot\partial_{X_{1}}\,,\quad
	\bar Y_{3}:=\tfrac12\,\partial_{U_{3}}\!\!\cdot(\partial_{X_{1}}-\partial_{X_{2}})\,,
	\label{bar Y}
\ee
which are nothing but the integration-by-parts versions of the $Y$-variables
(up to divergence terms, which we disregard). 
Notice as well that we have also redefined $\bar Y_{3}$ 
in order to get more symmetric expressions.}
As a result, the coupling $C(Y,Z)$ of eq.~\eqref{massless sol}
can be written, up to integrations by parts, as a new function $\bar C(\bar Y,Z)$ in different variables:
\be\label{bar coupling}
	C(Y,Z) \,\ett\,
	\bar C(\bar Y,Z)=e^{\l
	\left(Z_1\,\partial_{\bar Y_2}+Z_2\,\partial_{\bar Y_1}+
	Z_1\,Z_2\,\partial_{G}\right)\partial_{\bar Y_{3}}}\,
	K(\bar Y,G)\,\Big|_{G=G(\bar Y,Z)}\,,
\ee
where we have taken into account the contributions coming 
from the delta function in the measure.
The difference between $\bar C(Y,Z)$ and $C(Y,Z)$ lies in
the operators appearing in the exponent --- the former has only three terms while the latter \eqref{D op} has seven terms.
In particular, in flat space, they simply coincide with each other.

Given all this, we can express the gauge variation \eqref{f var tot} as
\be
	\delta^{\sst (0)}_{\sst E_{1}}\,S^{\sst (3)} \ett 
	\int_{\rm\sst (A)dS}
	e^{\partial_{U_{1}}\!\cdot\,\partial_{U_{2}}}\Big[
	T_{12}(X,U_{1})\,\partial_{X}^{\,2}\,\Phi_{3}(X,U_{2})
	\,+\,T_{13}(X,U_{1})\,\partial_{X}^{\,2}\,\Phi_{2}(X,U_{2})
	\Big]_{\sst U_{i}=0}\,,
	\label{1st deform}
\ee
where $T_{12}(X,U)$ is given by
\be
	T_{12}(X,U)
	=\frac12\,\partial_{Y_{1}}\bar C(Y, Z)\,E_{1}(X_{1},U_{1})\,
	\Phi_{2}(X_{2},U_{2})\,\Big|_{{}^{X_{i}=X}_{U_{i}=0}},
	\label{T12}
\ee
in terms of the \emph{newly defined} variables $Y$ and $Z$
(for simplicity of notations):
\ba \label{new YZ}
	& Y_{1}=\partial_{U_{1}}\!\!\cdot\partial_{X_{2}}\,,\quad
	Y_{2}=-\,\partial_{U_{2}}\!\!\cdot\partial_{X_{1}}\,,\quad
	Y_{3}=\tfrac12\,U\cdot(\partial_{X_{1}}-\partial_{X_{2}})\,,&\nn
	& Z_{1}=U\cdot\partial_{U_{2}}\,,\qquad\
	Z_{2}=U\cdot\partial_{U_{1}}\,,\qquad\
	Z_{3}=\partial_{U_{1}}\!\!\cdot\partial_{U_{2}}\,.&
\ea
Finally, comparing eq.~\eqref{1st deform} with eq.~\eqref{offshell cond}, 
we obtain the first-order deformation of the gauge transformations as
\be
	\delta^{\sst (1)}_{\sst E_{1}}\,\Phi_{3}(X,U)
	\ett -\,
	T_{12}(X,U)
	+U^{2}\,\a(X,U)+U\cdot X\,\b(X,U)+U\cdot\partial_{X}\,\g(X,U)\,.
	\label{delta 1}
\ee
Several comments are in order.
First, the $\ett$ above is to be considered with respect to the generating functions $E_{1}$ and $\Phi_{2}$
and it is inherited from the analogous equivalence relation on $\Phi_{1}$ and $\Phi_{2}$ in \eqref{1st deform};
Second, $\a$ and $\g$ terms are in the kernel
of the TT projection and so are inherited from the $\Phi_{3}$ part of the equality modulo traces and divergences\,; 
Lastly, the $\b$ term is introduced since the field 3 is tangent.
Hence, simplifying the problem by restricting our attention to the TT part
seems to leave too many ambiguities in the determination
of $\delta^{\sst (1)}_{\sst E_{1}}\Phi_{3}$\,, but actually, as we will see, 
these ambiguities can be fixed by general consistencies.
Let us make this point clear:
\begin{itemize}
\item First, the $\alpha$ term is what would be completely fixed if we had
worked with the full vertices instead of their TT parts.
Hence, in principle, its determination requires the information about the
trace and divergent pieces of the interactions.
However, without relying on an explicit treatment of such pieces,
the corresponding gauge-algebra deformations can be completely determined from the main piece $T_{ij}$
by asking consistency with the traceless conditions on gauge parameters
(in the Fronsdal setting).

\item The $\b$ term is a peculiarity of (A)dS couplings. 
Since any non-linear field theory in (A)dS can be reformulated
in the ambient-space formalism, the gauge transformations
of such a theory admit expressions as tangent ambient-space tensors. 
Hence, by requiring that
$\delta^{\sst (1)}_{\sst E_{1}}\Phi_{3}$ be tangent, one can fix $\b$\,.

\item The last term given by $\g$ corresponds to the genuine ambiguity 
related to redefinitions of gauge parameters.

\end{itemize}
Hence, modulo the genuine ambiguity $\g$ which we shall analyze later,
the deformation of gauge transformations can be written as
\be
	\delta^{\sst (1)}_{\sst E_{1}}\,\Phi_{3}(X,U)
	\ett -\,
	\Pi_{\Phi}\,T_{12}(X,U)\,,
	\label{delta 1'}
\ee
where $\Pi_{\Phi}$ is the projectors which make 
the corresponding bracket traceless (by choosing a suitable $\alpha$)
and renders the expression tangent (by choosing
a suitable $\beta$) for (A)dS cases. 
Moreover, we also assume that 
$\Pi_{\Phi}$ introduces in (A)dS a suitable power of $X^{2}/L^{2}$ as a factor 
(which is in fact another ambiguity not listed above)
so that the $\delta^{\sst (1)}_{\sst E_{1}}\Phi_{3}$ satisfies the homogeneity condition
as well.
Our prescription for $\Pi_{\Phi}$ may look a bit formal at this stage, but soon it will become clear that in this way we have access to many interesting informations on HS interactions.

The expression \eqref{delta 1'} for the first-order deformation of gauge transformations
is never vanishing for any of the cubic interactions \eqref{Cn}.
However, as anticipated, some of these deformations are trivial
since they can be reabsorbed by a gauge-field  redefinition.

\subsubsection*{Redefinitions of gauge fields and parameters} 

In order to analyze 
the effects of gauge-field and gauge-parameter redefinitions,
let us first express them in our framework.
Without loss of generality, we concentrate on redefinitions of $\Phi_{3}$ and $E_{3}$\,, which are given in terms of functions 
$\Omega_{3}\,, \Delta_{12}$ and $\Delta_{21}$ as
\ba
	&& \Phi_{3}(X,U)\ \to\
	\Phi_{3}(X,U)+\Pi_{\Phi}\,
	\Omega_{3}(A,B,Y,Z)\,
	\Phi_{1}(X_{1},U_{1})\,\Phi_{2}(X_{2},U_{2})\,
	\Big|_{{}^{X_{i}=X}_{U_{i}=0}}\,,\nn
	&& E_{3}(X,U)\ \to\
	E_{3}(X,U)+\Pi_{E}\,
	\D_{12}(A,B,Y,Z)\,
	E_{1}(X_{1},U_{1})\,\Phi_{2}(X_{2},U_{2})\,
	\Big|_{{}^{X_{i}=X}_{U_{i}=0}}\nn
	&& \hspace{116pt}
	+\, \Pi_{E}\,
	\D_{21}(A,B,Y,Z)\,
	E_{2}(X_{1},U_{1})\,\Phi_{1}(X_{2},U_{2})\,
	\Big|_{{}^{X_{i}=X}_{U_{i}=0}}\,.
	\label{redef}
\ea
Here, we introduced the operators $A$ and $B$ defined by
\be
	A=\tfrac12\,U\cdot\partial_{X}=\tfrac12\,U\cdot(\partial_{X_{1}}+\partial_{X_{2}})\,,
	\qquad
	B=\partial_{X_{1}}\!\!\cdot\partial_{X_{2}}
	\,.
\ee
As before, we introduce as well the projectors $\Pi_{\Phi}$
and $\Pi_{E}$ in order to make the redefinitions compatible 
with traceless and tangent gauge parameters.
The projector $\Pi_{E}$ has exactly the same role as $\Pi_{\Phi}$,
except that it adjusts the degree of homogeneity to match
that of gauge parameters. 

\subsection{Non-deforming couplings}

We are now ready to examine
how the redefinitions \eqref{redef} contribute to the first-order deformations of the gauge-transformations. 
When taking into account such contributions,
$\delta^{\sst (1)}_{\sst E_{1}}\Phi_{3}$ takes the following form:
\be
	\delta^{\sst (1)}_{\sst E_{1}}\,\Phi_{3}
	\ett \Pi_{\Phi}\left(-\,\tfrac12\,\partial_{Y_{1}}
	\bar C+
	[\,\O_{3}\,,\,U_{1}\cdot\partial_{X_{1}}]
	+U\cdot\partial_{X}\,\D_{12}
	\right) E_{1}\,\Phi_{2}\,
	\big|_{{}^{X_{i}=X}_{U_{i}=0}}\,,
	\label{deform delta}
\ee
where, using the chain rule, the commutator can be written as 
\be
	[\,\O_{3}\,,\, U_{1}\cdot\partial_{X_{1}}]
	=\left(A\,\partial_{Z_{2}}+B\,\partial_{Y_{1}}
	+Y_{3}\,\partial_{Z_{2}}-Y_{2}\,\partial_{Z_{3}}\right)\O_{3}\,.
\ee
Hence, for a given cubic interaction $\bar C$\,, the corresponding 
\mt{\delta^{\sst (1)}_{\sst E_{1}}\,\Phi_{3}} vanishes
if and only if there exist functions $\O_{3}$ and $\Delta_{12}$  
such that 
\ba
	&&-\,\tfrac12\,\partial_{Y_{1}}
	\bar C(Y,Z)+\left(A\,\partial_{Z_{2}}+B\,\partial_{Y_{1}}
	+Y_{3}\,\partial_{Z_{2}}-Y_{2}\,\partial_{Z_{3}}\right)
	\O_{3}(A,B,Y,Z)\nn
	&&\hspace{75pt}
	+\,2\,A\,\Delta_{12}(A,B,Y,Z)=0\,.
\ea
The $A$-dependence can be trivially solved by
taking $\D_{12}=-\partial_{Z_{2}}\O_{3}/2$ and 
assuming $\O_{3}$ independent of $A$\,.
Then the equation simplifies into 
\be
	-\,\tfrac12\,\partial_{Y_{1}}
	\bar C(Y,Z)+
	\left(B\,\partial_{Y_{1}}
	+Y_{3}\,\partial_{Z_{2}}-Y_{2}\,\partial_{Z_{3}}\right)
	\O_{3}(B,Y,Z)=0\,.
	\label{trivial delta}
\ee
In principle, for each cubic-interaction vertex: 
\be
	\bar P^{\sst [n]}_{s_{1}s_{2}s_{3}}(Y,Z):=
	e^{\l
	\left(Z_1\partial_{Y_2}+Z_2\partial_{Y_1}+
	Z_1Z_2\partial_{G}\right)\partial_{Y_{3}}}\,Y_{1}^{\,s_{1}-n}
	Y_{2}^{\,s_{2}-n} Y_{3}^{\,s_{3}-n} G^{n}\,\Big|_{G=G(Y,Z)},
	\label{bar Pn}
\ee
one can examine
whether there exists $\Omega_{3}$ satisfying eq.~\eqref{trivial delta} or not.

\paragraph{Flat-space case}

In flat space, eq.~\eqref{trivial delta} can be written as
\be
	\left(B\,\partial_{Y_{1}}
	+Y_{3}\,\partial_{Z_{2}}-Y_{2}\,\partial_{Z_{3}}\right)
	\big[\,C(Y,Z)+2\,B\,\Omega(B,Y,Z)\,\big]=0\,,
	\label{B eq}
\ee
using the fact that  
\mt{C=\bar C} satisfies 
\mt{(Y_{i}\,\partial_{Z_{j}}-Y_{j}\,\partial_{Z_{i}})\,C=0}\,.
The solution is given as an arbitrary polynomial
in six variables:
\be
	C+2\,B\,\Omega
	=\sum_{l_{i},m_{i},n_{i}} g_{\, l_{2},l_{3},m_{2},m_{3},n_{2},n_{3}}\,
	Y_{2}^{\, l_2}\, Y_{3}^{\, l_{3}}\, h_{12}^{\ m_{2}}\, h_{13}^{\ m_{3}}\,
	f_{2}^{\, n_{2}}\,f_{3}^{\, n_{3}}\,,
	\label{flat ND C}
\ee
where the $h_{1i}$'s and the $f_{i}$'s are given by
\ba
	& h_{12}(B,Y,Z):=-Y_{1}\,Y_{2}-B\,Z_{3}\,,
	\qquad
	& f_{2}(B,Y,Z):=Y_{2}\,G(Y,Z)+B\,Z_{1}\,Z_{3}\,,\nn
	&h_{13}(B,Y,Z):=-Y_{1}\,Y_{3}+B\,Z_{2}\,,
	\qquad
	&f_{3}(B,Y,Z):=-Y_{3}\,G(Y,Z)+B\,Z_{1}\,Z_{2}\,.
	\label{KB}
\ea
One can notice that $h_{1i}\approx H_{1i}$ up to integrations by parts.
Moreover, the $f_{i}$'s can be also related to $H$-like objects,
$h^{\pm}_{23}:=-Y_{2}\,Y_{3}\pm B\,Z_{1}$\,, as
\be	
	B\,f_{2}=Y_2^2\, h_{13}+h_{12}\,h_{23}^-\,,\qquad 
	B\,f_{3}=Y_3^2\,h_{12}+h_{13}\,h_{23}^+\,.
\ee
In fact, a subset --- but not all --- of the couplings \eqref{flat ND C} can be recast into
$H$-couplings of the form:
\be
	C(Y,Z)\approx
	\tilde K(Y_{\ell}, H_{12}, H_{23}, H_{31})
	\qquad [\ell=2\ \rm{or}\ 3]\,.
	\label{K H}
\ee
In order to see this point more clearly, let us restrict our attention to
the $s_{1}\!-\!s_{2}\!-\!s_{3}$ interactions.
Then, any consistent coupling is a linear combination of 
\mt{Y_{1}^{s_{1}-n}\,Y_{2}^{s_{2}-n}\,Y_{3}^{s_{3}-n}\,G^{n}}
with $n=0,\ldots,s_{\sst \rm min}$\,.
On the other hand, the solution \eqref{flat ND C} tells us
that any coupling with $\delta^{\sst (1)}_{\sst E_{1}}=0$ is a linear combination of
\be
	Y_{2}^{\,l_{2}}\,Y_{3}^{\,l_{3}}\,(Y_{1}\,Y_{2})^{m_{2}}\,(Y_{1}\,Y_{3})^{m_{3}}\,
	(Y_{2}\,G)^{n_{2}}\,(Y_{3}\,G)^{n_{3}}
	=Y_{1}^{m_{2}+m_{3}}\,Y_{2}^{l_{2}+m_{2}+n_{2}}\,Y_{3}^{l_{3}+m_{3}+n_{3}}
	\,G^{n_{2}+n_{3}},
\ee
with non-negative integers $l_{i}, m_{i}, n_{i}$'s  satisfying
\be
	s_{1}=m_{2}+m_{3}+n_{2}+n_{3}\,,\quad
	s_{2}=l_{2}+m_{2}+2\,n_{2}+n_{3}\,,\quad
	s_{3}=l_{3}+m_{3}+n_{2}+2\,n_{3}\,.
\ee
Analyzing the above inequalities, one can conclude that,
among all possible values of $n=0,\ldots,s_{\rm\sst min}$\,,
the  non-deforming couplings with $\delta^{\sst (1)}_{\sst E_{1}}=0$ correspond to
those satisfying
\be
	n\leq\tfrac{s_2+s_3-s_1}2\,.
	\label{Abel ineq}
\ee 
Equivalently, one can conclude that the number of derivatives, $\#_{\partial}$\,, of 
such couplings satisfies
\be
	\big(\,\#_{\partial}\ \textrm{in a flat-space coupling with}\ \delta^{\sst (1)}_{\sst E_{1}}=0
	\,\big)\ge 2\,s_{1}\,.
\ee
Now, let us move to the $H$-couplings given by eq.~\eqref{K H}.
The analysis carried out in Section \ref{sec: H} shows that 
they correspond to the couplings with
\be
	n\leq\tfrac{s_1+s_{2}+s_{3}}2-{\rm max}\{s_{1}, s_{\rm\sst mid}\}\,.
	\label{H ineq}
\ee
Hence, we have two distinct cases,
$s_{1}\neq s_{\rm\sst min}$ and $s_{1}=s_{\rm\sst min}$\,:

\begin{itemize}
\item
\textbf{$\bm{s_{1}\neq s_{\rm\sst min}}$\,:}
The two inequalities \eqref{Abel ineq} and \eqref{H ineq} coincide with each other.
Hence, any coupling
with $\delta^{\sst (1)}_{\sst E_{1}}=0$ can be written as 
a $H$-coupling \eqref{K H};

\item 
\textbf{$\bm{s_{1}= s_{\rm\sst min}}$\,:}
In this case, there exist couplings in the range:
\be
	\tfrac{s_{\rm\sst max}+s_{\rm\sst min}-s_{\rm\sst mid}}2<
	n\leq\tfrac{s_2+s_3-s_1}2\,,
	\label{L ineq}
 \ee
 which neither deform $\delta^{\sst (1)}_{\sst E_{1}}$ nor are
 expressible as a $H$-coupling \eqref{K H}.

\end{itemize}

\paragraph{(A)dS case}

For the analysis of (A)dS cubic interactions, 
we need to come back to the original condition \eqref{trivial delta}.
The basic way of solving the problem is Taylor expanding $\O_{3}$ as
\be
	\Omega_{3}(B,Y,Z)=\sum_{m=0}^{\infty}\,
	B^{m}\,\Omega^{\sst [m]}(Y,Z)\,,
\ee
and recasting  \eqref{trivial delta} into a set of equations:
\be
	\partial_{Y_{1}}\,\Omega^{\sst [m]}(Y,Z)
	=(Y_{2}\,\partial_{Z_{3}}-Y_{3}\,\partial_{Z_{2}})\,
	\Omega^{\sst [m-1]}(Y,Z)\,.
	\label{recur cond}
\ee
For a given $\O^{\sst [-1]}=\frac12\,\bar C(Y,Z)$\,, 
we recursively solve for $\O^{\sst [m]}$'s.
If there is no obstruction in solving \eqref{recur cond} to all order, then 
the corresponding $\bar C(Y,Z)$ does not induce any non-trivial deformation 
of the gauge transformations.
For this analysis, let us first consider $\O^{\sst [-1]}$ given 
by a monomial: 
\be
	\Omega^{\sst [-1]}(Y,Z)
	=Y_{1}^{\s_{1}}\,Y_{2}^{\s_{2}}\,Y_{3}^{\s_{3}}\,
	Z_{1}^{\t_{1}}\,Z_{2}^{\t_{2}}\,Z_{3}^{\t_{3}}\,.
	\label{mono C}
\ee
Then, one can show that the equations \eqref{recur cond} admit a solution if (see Appendix~\ref{Appendix: B}):
\be
	\s_{2}-\t_{2}\ge \s_{1}\qquad {\rm or} \qquad 
	\s_{3}-\t_{3}\ge \s_{1}\,.
	\label{generic bound}
\ee
However, depending on linear combinations of monomials \eqref{mono C},
a solution may exist for a wider class of polynomials $\O^{\sst [-1]}$, although
each monomial term does not admit a solution independently.
For instance,
for the flat-space couplings, the bound is given in \eqref{Abel ineq} which
is equivalent to
\be
	\s_{2}-\t_{2}+\s_{3}-\t_{3}\ge \s_{1}\,.
	\label{flat bound}
\ee
One can show that this is weaker than \eqref{generic bound},
although it is not manifest.
Having this in mind, let us consider 
the case $\bar C=\bar P^{\sst [n]}_{s_{1}s_{2}s_{3}}$ \eqref{bar Pn}\,. 
One can first notice that 
$\O^{\sst [-1]}$ can be expanded as a power series in 
$\l$\,, and examined independently for each of the expansion coefficients.
Hence, we can focus on
\be
	\Omega^{\sst [-1]}(Y,Z)
	=\left(Z_1\partial_{Y_2}+Z_2\partial_{Y_1}+
	Z_1Z_2\partial_{G}\right)^{k} \partial_{Y_{3}}^{\ k}\
	Y_{1}^{\,s_{1}-n}
	Y_{2}^{\,s_{2}-n} Y_{3}^{\,s_{3}-n} G^{n}\Big|_{G=G(Y,Z)}\,.
	\label{O-1}
\ee
Each term in the above function, as the monomials \eqref{mono C}, satisfies
\be
	\s_{1}=s_{1}\,,\qquad
	\s_{2}-\t_{2}=s_{2}-n-k\,,\qquad
	\s_{3}-\t_{3}=s_{3}-n-k\,,
\ee
and should satisfy \eqref{generic bound} in generic cases
or a slightly weaker condition
as in the case of \eqref{flat bound}.
The precise analysis --- whether the differential operator 
in \eqref{O-1} generates a particular linear combination which extends
 the bound \eqref{generic bound} --- is tedious and non-trivial,
so let us draw some general lessons.
If the flat-space limit of a (A)dS coupling induces a non-trivial deformation,
then it also does in (A)dS: this allows us to examine 
$\bar C=\bar P^{\sst [n]}_{s_{1}s_{2}s_{3}}$ 
only for $n\ge (s_{2}+s_{3}-s_{1})/2$ for the identification of non-deforming couplings.
However, the inverse is not true in general:
even for the couplings whose flat-space limit
satisfies \eqref{flat bound}, and hence does not lead to any non-trivial deformation,
their $\Lambda^{k}$ corrections corresponding to \eqref{O-1}
can induce a non-trivial $\delta^{\sst (1)}$ for a large enough $k$\,.
The reason is that the two quantities $\s_{2}-\t_{2}$ and  $\s_{3}-\t_{3}$ 
appearing in the generic bound \eqref{generic bound} and also in the 
special bound \eqref{flat bound}
both decrease as $k$ increases.
Hence, there must exist a critical value $k=k^{c}_{1}(s_{1},s_{2},s_{3},n)$\,, 
where the equation \eqref{recur cond} admits no solution,
and the deformation of gauge transformation 
induced by the coupling $\bar C=\bar P^{\sst [n]}_{s_{1}s_{2}s_{3}}$ reads
\be
	\delta^{\sst (1)}_{\sst E_{1}}\Phi_{3}
	=\cO\!\left(\Lambda^{k^{c}_{1}(s_{1},s_{2},s_{3},n)}\right).
	\label{delta 1 AdS}
\ee
Although we do not identify $k^{c}_{1}(s_{1},s_{2},s_{3},n)$ precisely,
we can have some information on it. 
For instance, following the generic bound \eqref{generic bound}, 
we can conclude that $k^{c}_{1}$ satisfies
\be
	k^{c}_{1} > {\rm max}\{s_{2},s_{3}\}-s_{1}-n\,,
\ee
and in the particular case of the minimum-derivative $2\!-\!s\!-\!s$ interactions, it is
\be
	k^{c}_{1} =s-2\,,
\ee
by compatibility with the Fradkin-Vasiliev results \cite{Fradkin:1986qy,Fradkin:1987ks}.
The equation \eqref{delta 1 AdS} may give us a wrong impression that in (A)dS
all couplings induce non-trivial deformations. It is because
we have considered the deformations of a single coupling $\bar C=\bar P^{\sst [n]}_{s_{1}s_{2}s_{3}}$ alone.
For the analysis of deformations, in flat space, it was sufficient to 
examine each monomial $\bar P^{\sst [n]}_{s_{1}s_{2}s_{3}}$
since different couplings never talk to each other having different number of 
derivatives.
On the contrary, this is no more true in (A)dS, so we need to consider general linear
combinations with a fixed number of highest derivatives:
\be
	\bar Q^{\sst [n]}_{s_{1}s_{2}s_{3}}
	=\bar P^{\sst [n]}_{s_{1}s_{2}s_{3}}
	+c_{1}\,\l\,\bar P^{\sst [n+1]}_{s_{1}s_{2}s_{3}}
	+c_{2}\,\l^{2}\,\bar P^{\sst [n+2]}_{s_{1}s_{2}s_{3}}
	+\cdots \qquad [c_{1}, c_{2}, \ldots \in \mathbb R]\,.
	\label{lin comb C}
\ee
Hence,
there may exist more than two terms in the above  
giving the same leading $\L$-order deformations. 
In such cases, those deformations might be cancelled among each others.
In that regard,
let us note that, 
as in the flat-space case, one can choose a basis of couplings such that the ones satisfying the condition \eqref{H ineq} are expressed 
as $H$-couplings \eqref{K H} and hence do not induce any deformation $\delta^{\sst (1)}_{\sst E_{1}}$\,.
In other words, starting from the $H$-couplings,
one can derive the proper linear combinations 
$\bar Q^{\sst [n]}_{s_{1}s_{2}s_{3}}$ \eqref{lin comb C} leading to 
precise cancellation of $\delta^{\sst (1)}_{\sst E_{1}}$
(for more detailed analysis, see Section \ref{sec: H}). 
Hence, one can summarize the result as follows:
\begin{itemize}
\item 
\textbf{$\bm{s_{1}\neq s_{\rm\sst min}}$\,:}
 For $n\le \frac{s_{2}+s_{3}-s_{1}}2$\,,
 one can find the couplings $\bar Q^{\sst [n]}_{s_{1}s_{2}s_{3}}$
 which are 
related to $H$-couplings, 
so they are manifestly all non-deforming;
\item
\textbf{$\bm{s_{1}= s_{\rm\sst min}}$\,:}
For $n$ satisfying \eqref{L ineq},
the corresponding gauge-transformation deformation
is generically non-trivial.
However, one needs in principle to examine all possible $\bar Q^{\sst [n]}_{s_{1}s_{2}s_{3}}$'s
in order to 
see whether there might be a further cancellation leading to $\delta^{\sst (1)}_{\sst E_{1}}=0$\,.
Let us note first that when $s_{1}\le  \frac{s_{2}+s_{3}-s_{1}}2$\,,
the minimum-derivative coupling with $n=s_{1}$ also belongs to this class,
and it is clear for this coupling that the deformation 
cannot be cancelled since there is no other term in \eqref{lin comb C}.
More generally, one can conclude that for the latter couplings this cancellation never happens 
because otherwise it would contradict the compatibility with the non-triviality of the global-symmetry deformations (see Section \ref{sec: Global}).

\end{itemize}

\subsubsection*{Remarks}

The results obtained in this section are summarized in Table~\ref{tab: classification}. 
Let us conclude with a few more remarks:
\begin{itemize}
\item
One can see that \mt{\delta^{\sst (1)}_{\sst E_{i}}\Phi_{i+1}=0} implies 
\mt{\delta^{\sst (1)}_{\sst E_{i}}\Phi_{i-1}=0} and vice versa. 
Hence, when we discuss whether 
the deformations are vanishing or not,
we can refer them to $\delta^{\sst (1)}_{\sst E_{i}}$ without writing explicitly on which field they act upon.
\item
The $H$-couplings \eqref{K H} are non-deforming since 
$Y_{2}, Y_{3}$ and $H_{ij}$'s commute with
the free gauge transformation $\delta^{\sst (0)}_{\sst E_{1}}\Phi_{3}$
without relying on the EoM, irrespectively of the curvature of spacetime.
However, if we allow the dependence of both $Y_{2}$ and $Y_{3}$
at the same time
in the coupling \eqref{K H},
then it does not satisfy the (A)dS gauge invariance condition any more, so it is important to allow only one between $Y_{2}$ and $Y_{3}$\,. 
\item
In flat space, there is a class of couplings with $s_{1}=s_{\rm\sst min}$ satisfying 
the condition \eqref{L ineq}  
for which only two of the three deformations are non-vanishing:
\be
	\delta^{\sst (1)}_{\sst E_{1}}=0\,,\qquad
	\delta^{\sst (1)}_{\sst E_{2,3}}\neq 0\,.
	\label{non cyclic delta 1}
\ee
The corresponding (A)dS couplings deform all three gauge transformations:
\be
	\delta^{\sst (1)}_{\sst E_{1}}=
	\cO\big(\Lambda^{k^{c}_{1}}\big)\,,\qquad
	\delta^{\sst (1)}_{\sst E_{2,3}}=\mathcal O\big(\Lambda^{0}\big)\neq0\,,
\ee
due to the presence of lower-derivative terms in the coupling. 

\end{itemize}

\subsubsection*{Examples:
Gravitational interactions of spin-three and -four fields}

Let us conclude this section with some concrete examples:
the cubic vertices for the \mt{3\!-\!3\!-\!2} and \mt{4\!-\!4\!-\!2} interactions.
These are of particular interest since they show
how HS fields (spin 3 and spin 4) can gravitationally interact with
spin 2 in a (A)dS background.
In each cases, there exist  three consistent vertices, while
only the ones with the minimum number of highest-derivatives
(four for spin 3 and six for spin 4):
\be
	K^{\sst 3\!-\!3\!-\!2}(Y,G)=\frac k2\,Y_{1}\,Y_{2}\,G^{2}\,,\qquad
	K^{\sst 4\!-\!4\!-\!2}(Y,G)=\frac{k'}8\,Y_{1}^{2}\,Y_{2}^{2}\,G^{2}\,,
	\label{gravi coup}
\ee
induce non-trivial deformations of gauge transformations.
Following the analysis presented in this section,
one can compute the corresponding $\bar C$'s by
applying $e^{\lambda\left(Z_1\partial_{Y_2}+Z_2\partial_{Y_1}+
	Z_1Z_2\partial_{G}\right)\partial_{Y_{3}}}$
and then extract the gauge-transformation deformation $\delta^{\sst (1)}$
according to \eqref{deform delta}.
With proper gauge-parameter redefinitions, controlled by $\D_{ij}$'s,
one can remove the higher-derivative contributions in $\delta^{\sst (1)}$\,.
All in all, 
under the spin-2 gauge transformation with parameter $\Xi$\,,
the spin-3 field $\Phi$ transforms as 
\ba
	\delta^{\sst (1)}_{\Xi}\Phi(X,U)
	&\!\ett\!& -3\,k\,\l
	\left(\tfrac1{3!}\,Y_{1}\,Z_{1}^{3}+
	\tfrac12\,Y_{2}\,Z_{1}^{2}\,Z_{2}\right)
	\Xi(X_{1},U_{1})\,\Phi(X_{2},U_{2})\,
	\Big|_{{}^{X_{i}=X}_{U_{i}=0}}\nn
	\eq
	-3\,k\,\l\left(\partial_{U_{1}}\!\cdot\partial_{X_{2}}
	-\partial_{X_{1}}\!\cdot\partial_{U_{2}}\right)
	\Xi(X_{1},U_{1})\,\Phi(X_{2},U_{2})\,
	\Big|_{{}^{X_{i}=X}_{U_{i}=U}}\,,
	\label{332 tr}
\ea
and the spin-4 field $\Psi$ as
\ba
	\delta^{\sst (1)}_{\Xi}\Psi(X,U)
	&\!\ett\!& -6\,k'\,\l^{2}
	\left(\tfrac1{4!}\,Y_{1}\,Z_{1}^{4}+
	\tfrac1{3!}\,Y_{2}\,Z_{1}^{3}\,Z_{2}\right)
	\Xi(X_{1},U_{1})\,\Psi(X_{2},U_{2})\,
	\Big|_{{}^{X_{i}=X}_{U_{i}=0}}\nn
	\eq
	-6\,k'\,\l^{2}\left(\partial_{U_{1}}\!\cdot\partial_{X_{2}}
	-\partial_{X_{1}}\!\cdot\partial_{U_{2}}\right)
	\Xi(X_{1},U_{1})\,\Psi(X_{2},U_{2})\,
	\Big|_{{}^{X_{i}=X}_{U_{i}=U}}\,.
	\label{442 tr}
\ea
Analogous expressions can be easily extracted in the general case
of $s\!-\!s\!-\!2$ interactions.
As one can notice, these transformations 
are given by Lie derivatives and correspond 
to nothing but the general coordinate transformations
of tensor fields $\Phi$ and $\Psi$\,.
Hence, one can interpret these interactions as gravitational ones.
Notice also that the transformations
are proportional to $\l\sim \L$ or $\l^{2}\sim\L^{2}$\,,
hence they vanish in the flat-space limit.
It is also interesting to note that the 
gravitational interactions \eqref{gravi coup} of HS fields
induce deformations 
also for the graviton field $\G$\,:
under the spin-3 gauge transformation with parameter $E$\,, it transforms as
\ba
	&&\delta^{\sst (1)}_{E}\,\G(X,U)
	\ett
	-3\,k\,B\,Z_{3}
	\left(\tfrac12\,Y_{1}\,Z_1^2 +
	Y_{2}\,Z_1\,Z_{2}\right) 
	E(X_{1},U_{1})\,\Phi(X_{2},U_{2})\,
	\Big|_{{}^{X_{i}=X}_{U_{i}=0}}\nn
	&&=
	-3\,k\left(\partial_{U_{1}}\!\cdot\partial_{X_{2}}
	-\partial_{X_{1}}\!\cdot\partial_{U_{2}}\right)
	\partial_{X_{1}}\!\cdot\partial_{X_{2}}\,
	\partial_{U_{1}}\!\cdot\partial_{U_{2}}\,
	E(X_{1},U_{1})\,\Phi(X_{2},U_{2})\,
	\Big|_{{}^{X_{i}=X}_{U_{i}=U}}\,,\quad
	\label{332 tr'}
\ea
while under the spin-4 gauge transformation with parameter $F$\,, it transforms as
\ba
	&&\delta^{\sst (1)}_{F}\,\G(X,U)\ett
	-\tfrac32\,k'\,B^{2}\,Z_{3}^{2}
	\left(\tfrac12\,Y_{1}\,Z_1^2 +
	Y_{2}\,Z_1\,Z_{2}\right) 
	F(X_{1},U_{1})\,\Psi(X_{2},U_{2})\,
	\Big|_{{}^{X_{i}=X}_{U_{i}=0}}\nn
	&&=
	-6\,k'\left(\partial_{U_{1}}\!\cdot\partial_{X_{2}}
	-\partial_{X_{1}}\!\cdot\partial_{U_{2}}\right)
	\frac{(\partial_{X_{1}}\!\cdot\partial_{X_{2}})^{2}}2
	\frac{(\partial_{U_{1}}\!\cdot\partial_{U_{2}})^{2}}2
	F(X_{1},U_{1})\,\Psi(X_{2},U_{2})\,
	\Big|_{{}^{X_{i}=X}_{U_{i}=U}}\,.\qquad
	\label{442 tr'}
\ea
Such transformations of the spin-2 field do not involve 
$\l$ dependence, they survive in the flat-space limit.
Let us also remark that all these expressions
(\ref{332 tr}\,-\,\ref{442 tr'}) are tangent and 
homogeneous with the proper degrees.

\section{Deformations of gauge algebras and abelian couplings} 
\label{sec: gaugealgebra}

\subsection{Deformations of gauge algebras}

After obtaining the first order deformation of the
gauge transformations $\delta^{\sst (1)}_{\sst E_{i}}$\,, 
we can identify the lowest-order part of the gauge algebra.
For that, we need to compute
$\delta^{\sst (0)}_{\sst E_{1}}\,\delta^{\sst (1)}_{\sst E_{2}}$ and 
$\delta^{\sst (0)}_{\sst E_{2}}\,\delta^{\sst (1)}_{\sst E_{1}}$\,. 
We can restrict the attention to the latter since the former 
can be obtained from the latter by relabeling.
This is given by
\be\label{gauge comm}
	\delta^{\sst (0)}_{\sst E_{2}}\delta^{\sst (1)}_{\sst E_{1}}\,\Phi_{3}
	\ett \Pi_{\Phi}
	\left[-\,\tfrac12\partial_{Y_{1}}\bar C+2\,A\,\D_{12}\,,\,U_{2}\cdot\partial_{X_{2}}\,\right]
	E_{1}\,E_{2}\,
	\big|_{{}^{X_{i}=X}_{U_{i}=0}}\,,
\ee
where we have included the contribution of gauge-parameter redefinitions, but
omitted the contribution of gauge-field redefinitions, since they 
automatically drops out by taking the antisymmetric part 
to build the commutator.
The commutator of two gauge transformations readily takes the form of a free gauge transformation:
\be
	\delta^{\sst (0)}_{\sst [E_{2}}\,\delta^{\sst (1)}_{\sst E_{1}]}\,\Phi_{3}
	=
	U\cdot\partial_{X}\,[\![\,E_{2}\,,\,E_{1}\,]\!]^{\sst (0)}_{3}\,,
\ee
with the gauge parameter given by
\ba
	[\![\,E_{1}\,,\,E_{2}\,]\!]^{\sst (0)}_{3}&\!\ett\!&
	\tfrac12\,\Pi_{E}\,\big[
	\tfrac12\left(\partial_{Y_{1}}\partial_{Z_{1}}
	+\partial_{Y_{2}}\partial_{Z_{2}}\right)\bar C
	-\left(A\,\partial_{ Z_{1}}
	-B\,\partial_{Y_{2}}-Y_{3}\,\partial_{Z_{1}}+Y_{1}\,\partial_{Z_{3}}\right)
	\D_{12}\nn
	&& \hspace{40pt} +\left(A\,\partial_{ Z_{2}} 
	+B\,\partial_{Y_{1}}+Y_{3}\,\partial_{Z_{2}}-Y_{2}\,\partial_{Z_{3}}\right)
	\D_{21}\big]\,
	E_{1}\,E_{2}\,
	\big|_{{}^{X_{i}=X}_{U_{i}=0}}\,.
	\label{commutator}
\ea
Here we have used the fact that
\be
	\Pi_{\Phi}\,U\cdot \partial_{X}=U\cdot \partial_{X}\,\Pi_{E}\,.
\ee
Hence, the formula \eqref{commutator} provides the general expression for the 
lowest-order deformation of gauge algebras (modulo the terms proportional to divergences and d'Alembertians of gauge parameters).

\subsection{Abelian interactions}

We can now ask for a given cubic interaction $\bar C$ whether the 
resulting commutators are Abelian, that is \mt{[\![\,E_{1}\,,\,E_{2}\,]\!]^{\sst (0)}_{3}=0}\,,
up to a gauge-parameter redefinition.
This is equivalent to ask whether there exists a solution
$(\Delta_{12}, \D_{21})$ to 
\begin{multline}
	\tfrac12\left(\partial_{Y_{1}}\partial_{Z_{1}}
	+\partial_{Y_{2}}\partial_{Z_{2}}\right)\bar C(Y,Z)
	-\left(A\,\partial_{ Z_{1}}
	-B\,\partial_{Y_{2}}-Y_{3}\,\partial_{Z_{1}}+Y_{1}\,\partial_{Z_{3}}\right)
	\D_{12}(A,B,Y,Z)\\
	+\left(A\,\partial_{ Z_{2}} 
	+B\,\partial_{Y_{1}}+Y_{3}\,\partial_{Z_{2}}-Y_{2}\,\partial_{Z_{3}}\right)
	\D_{21}(A,B,Y,Z)=0\,.
	\label{trivial alg}
\end{multline}
The above equation has a similar structure to
the one \eqref{trivial delta} associated with $\delta^{\sst (1)}$,
but is more involved including this time two functions $\D_{12}$ and $\D_{21}$.
Hence, we take a different approach, instead of solving directly eq.~\eqref{trivial alg}\footnote{ 
With the help of Mathematica, we have analyzed all the cases up to spin $10$ checking the 
agreement with the classification obtained in the other way.
}, 
to address the issue of gauge-algebra deformation. 
For that, let us begin by considering the following proposition:
for a given cubic interaction $\bar C$\,,
the corresponding deformation of gauge algebra is trivial
if and only if at least one of two gauge-transformation deformations
$\delta^{\sst (1)}_{\sst E_{1}}$ and $\delta^{\sst (1)}_{\sst E_{2}}$ is trivial:
\be\label{observation}
	[\![\,E_{1}\,,\,E_{2}\,]\!]^{\sst (0)}_{3}=0
	\qquad
	\Longleftrightarrow \qquad
	\delta^{\sst (1)}_{\sst E_{1}}=0\quad{\rm or}\quad 
	\delta^{\sst (1)}_{\sst E_{2}}=0\,.
\ee
The implication $\Leftarrow$ is manifest: assuming  \mt{\delta^{\sst (1)}_{\sst E_{1}}=0}\,,
we get 
\mt{\delta^{\sst (0)}_{\sst [\![E_{1},E_{2}]\!]^{\sst (0)}_{3}}=\delta^{\sst (0)}_{\sst E_{1}}\,
\delta^{\sst (1)}_{\sst E_{2}}} which is precisely
the form of a gauge-parameter deformation.
The implication $\Rightarrow$ can be shown by 
checking that
the couplings with non-trivial global-symmetry
deformations --- which necessarily deform the gauge algebra ---
exhaust all couplings with  
$\delta^{\sst (1)}_{\sst E_{1}}, \delta^{\sst (1)}_{\sst E_{2}}\neq 0$\,. 
Leaving the issue of the global-symmetry deformations
to the next section,
let us conclude this section by presenting 
the classification separately for flat-space and (A)dS cases.
\paragraph{Flat-space case} Taking the union of the two conditions \eqref{Abel ineq},
one concludes that a coupling with the number of derivatives:
\be
	\#_{\partial} < 2\,\min\{s_{1},s_{2}\}\,,
\ee
leads to the deformation of algebra, \mt{[\![\,E_{1}\,,\,E_{2}\,]\!]^{\sst (0)}_{3}\neq 0}\,.

\paragraph{(A)dS case} In (A)dS, only $H$-couplings \eqref{K H} result in a trivial deformation of the gauge transformations.  
Hence, any coupling with  \mt{[\![ \,E_{1}\,,\,E_{2}\,]\!]_{3}^{\sst (0)}=0}
can be expressed as
\be
	C(Y,Z)\approx
	\tilde K(Y_{\ell}, H_{12}, H_{23}, H_{31})
	\qquad [\ell=1,2\ {\rm or}\ 3]\,,
	\label{Abelian}
\ee
which corresponds to the union between the $H$-couplings  \eqref{K H}
and their $\delta^{\sst (1)}_{\sst E_{2}}$ counterparts.
This is equivalent to say that a coupling with the number of derivatives:
\be
	\#_{\partial} < 2\,s_{\rm\sst mid}\,,
\ee
leads to a deformation of the algebra, \mt{[\![\,E_{1}\,,\,E_{2}\,]\!]^{\sst (0)}_{3}\neq 0}\,.

\section{Global symmetries}
\label{sec: Global}

\subsection{Killing tensors}

The settings for the analysis of gauge-symmetry deformations
can be also used to study the global (or rigid) symmetries of HS theory --- the HS algebras.
The latter can be obtained from the gauge symmetries 
by imposing the Killing equation:\footnote{See \cite{Bekaert:2005ka,Bekaert:2006us}
for previous works on HS killing tensors.}
\be
	U\cdot\partial_{X}\,\bar E(X,U)=0\,.
	\label{Killing}
\ee
Together with the traceless condition on the gauge parameter
and the homogeneity and tangentiality conditions \eqref{HT},\footnote{
The operators
$e=X\cdot \partial_{U}$, $f=U\cdot \partial_{X}$ and 
$h=X\cdot \partial_{X}-U\cdot\partial_{U}$
can be viewed as the generators of $\mathfrak{sp}_{2}$:
\be
	[h,e]=2\,e\,,\qquad [h,f]=-2\,f\,,\qquad [e,f]=h\,.
\ee
Then, the massless Killing tensors are $\mathfrak{sp}_{2}$ singlets
reflecting the construction of HS algebra based on Howe duality.
For the partially-massless cases with  $\mu\in \mathbb N$\,, 
the Killing tensors carry rather $(2\,\mu+1)$-dimensional representations:
\be
	e\,\bar E=0\,,\qquad f^{\mu+1}\,\bar E=0\,,\qquad
	(h-2\,\mu)\,\bar E=0\,.
\ee
On the other hand, the fields carry infinite dimensional 
representations of $\mathfrak{sp}_{2}$\,:
\be
	e\,\Phi=0\,,\quad (h-\mu)\,\Phi=0\,,\quad 
	\Phi\sim \Phi+f^{\mu+1}\,E\qquad
	[\,e\,E=0\,,\quad (h-2\,\mu)\,E=0\,]\,. 
\ee
}
the Killing equation \eqref{Killing} results in additional %$\mathfrak{sp}_{2}$ 
conditions:
\be
	\partial_{X}\!\cdot\partial_{U}\,\bar E(X,U)=0\,,\qquad
	\partial_{X}^{\,2}\,\bar E(X,U)=0\,,\qquad
	\partial_{U}^{\,2}\,\bar E(X,U)=0\,.
\ee
The solutions to these conditions for the parameters $\bar E$ --- namely Killing tensors --- forms a HS algebra.
A generic element of the latter:
\be
	\bar E(X,U)=\sum_{r=0}^{\infty}\,\frac{1}{(r!)^{2}}\,
	\bar E_{\sst M[r]}\,T^{\sst M[r]}\,,
	\label{decomp killing}
\ee
can be decomposed into the generators of the corresponding HS algebra as
\be
	T^{\sst M_{1}\cdots M_{r}\,M'_{1}\cdots M'_{r}}=
	X^{\sst  [M_{1}} U^{\sst M'_{1}]}
	\cdots\,X^{\sst [M_{r}}\,
	U^{\sst M'_{r}]}\,.
\ee
In eq.~\eqref{decomp killing}, we have used a short-hand
notation $\st M[r]$ for the superscript 
$\st  M_{1}\cdots M_{r}\,M'_{1}\cdots M'_{r}$\,.
Let us notice here that since the generators $T^{\sst M[r]}$
appear always contracted with the traceless parameter $\bar E_{\sst M[r]}$\,,
they are defined actually as an equivalence class:
\be
	T^{\sst M[r]} \sim T^{\sst M[r]}
	+X\cdot U\,P_{1}^{\sst M[r]}
	+X^{2}\,P_{2}^{\sst M[r]}+U^{2}\,P_{3}^{\sst M[r]}\,,
\ee
where $P_{i}^{\sst M[r]}$'s are arbitrary polynomial in $X$ and $U$ possibly involving the ambient metric tensor $\eta_{\sst MN}$ and compatible with the homogeneity degree of $T^{\sst M[r]}$\,. 
From the gauge-algebra brackets \eqref{commutator} given by \mt{s_{1}\!-\!s_{2}\!-\!s_{3}} cubic interactions, 
we can then deduce the \mt{r_{1}\!-\!r_{2}\!-\!r_{3}} part 
($r_{i}\equiv s_{i}-1$)
of the structure constant.

\subsection{Higher-spin algebras}

Now let us consider the bracket \eqref{commutator} where
the gauge parameters are Killing tensors.
The cubic consistency ensures that 
the bracket of two Killing tensors gives again a Killing tensor:
\be
	U\cdot\partial_{X}\,\big[\hspace{-3pt}\big[\,\bar E_{1}\,,\,
	\bar E_{2}\,
	\big]\hspace{-3pt}\big]^{\sst (0)}_{3}=0\,.
	\label{Killing BR}
\ee 
About the explicit form of the bracket, we have two major simplifications. 
First, the TT restriction does not exclude any term
since all TT parts vanish when the Killing equations hold.
Second, the reparameterization  terms given by $\D_{ij}$ also vanish
since they involve $U_{i}\cdot\partial_{X_{i}}\,\bar E_{i}$\,.
Hence, the bracket of the global-symmetry algebra:
\ba
	\big[\hspace{-3pt}\big[\,\bar E_{1}\,,\,
	\bar E_{2}\,
	\big]\hspace{-3pt}\big]^{\sst (0)}_{3}=
	\Pi_{E}\,\cF^{s_{3}}{}_{s_{1}s_{2}}(Y,Z)\, \bar E_{1}\, \bar E_{2}\,
	\big|_{{}^{X_{i}=X}_{U_{i}=0}}\,,
	\label{global}
\ea
 is given simply by 
\be
	\cF^{s_{3}}{}_{s_{1}s_{2}}(Y,Z)=\tfrac14
\left(\partial_{Y_{1}}\partial_{Z_{1}}
	+\partial_{Y_{2}}\partial_{Z_{2}}\right)\bar C(Y,Z)\,,
\ee
and there remains no term unfixed or ambiguous.
Herefrom, we suppress the superscript $(0)$ of the bracket
since it is the complete one for the global symmetries.
Incidentally, the absence of ambiguity at the global-symmetry level
is also true for $\delta^{\sst (1)}_{\sst \bar E_{i}}$\,.
Let us recall that $\bar C(Y,Z)$ is given explicitly through the coupling function 
$K$ which can be expanded for given spins \mt{s_{1}\!-\!s_{2}\!-\!s_{3}} 
as follows:
\be
	K(Y_{1},Y_{2},Y_{3},G)=
	\sum_{n=0}^{s_{\sst\rm min}}
	\,k_{n}\,
	\frac{Y_{1}^{s_{1}-n}}{(s_1-n)!}\, \frac{Y_{2}^{s_{2}-n}}{(s_2-n)!}\, \frac{Y_{3}^{s_{3}-n}}{(s_3-n)!}\, 
	\frac{G^{n}}{n!}\,.
	\label{K k}
\ee
In the following, we explore the implications of HS global symmetries separately
for (A)dS and for flat space.  

\subsubsection*{(A)dS}

The coupling $\bar C$ contains $K(Y,G)$ as its highest-derivative part, 
while the lower-derivative parts are generated by the operator $e^{\l\left(Z_1\,\partial_{Y_2}+Z_2\,\partial_{Y_1}+
	Z_1\,Z_2\,\partial_{G}\right)\partial_{Y_{3}}}$\,.
One can count the number
of derivatives involved in the coupling associated with $G^{n}$\,, and its minimum is given by
\be
	s_1+s_2+s_3-2n-2(s_3-n)=\,s_1+s_2-s_3\,,
\ee
independently on the initial number of derivatives of the coupling. 
This implies that the lowest-derivative part of the above commutator contains exactly the right number of derivatives to be compatible with the homogeneity conditions \eqref{HT}:
\ba
	\big(\,\D_{X}\textrm{ of }[\![E_{1} ,E_{2} ]\!]_{3}\,\big)\eq
	\big(\,\D_{X}\textrm{ of }E_{1}\,\big) + 
	\big(\,\D_{X}\textrm{ of }E_{2}\,\big) -
	\big(\,\#_{\partial}\textrm{ of }[\![ E_{1} , E_{2}]\!]_{3}\,\big)\nn
	\eq
	(s_{1}-1)+(s_{2}-1)-(s_{1}+s_{2}-s_{3}-1) \nn
	\eq s_{3}-1=\big(\,\D_{X}\textrm{ of }E_{3}\,\big)\,.
\ea
The higher-derivative terms have lower degree of homogeneity in $X$ and need to be supplemented by appropriate powers of $X^2$ accordingly to the projector $\Pi_E$ --- recall that the projector $\Pi_{E}$ is making the commutator traceless,
tangent and also of the proper homogeneity degree. Therefore, all the higher-derivative terms are trace components,
which are annihilated by the action of $\Pi_{E}$\,.
Hence, the lowest-derivative part of the cubic interaction $\bar C$
contains the full information on 
the structure constant of the global symmetries. 
The commutator can then be readily computed starting from the explicit formulas for the coupling $\bar C(Y,Z)$ and reads\footnote{To be more precise, $k_{n}$'s must involve totally antisymmetric structure constant as well 
if \mt{s_{1}+s_{2}+s_{3}} is an odd integer.}
\ba	
	&& \cF^{s_{3}}{}_{s_{1}s_{2}}(Y_{1},Y_{2},Z_{1},Z_{2},Y_{3}\,Z_{3})= \frac14\,
	\left(\partial_{Y_{1}}\partial_{Z_{1}}+\partial_{Y_{2}}\partial_{Z_{2}}\right)
	\sum_{n=0}^{s_{\rm\sst min}}\,k_{n}\,\l^{s_{3}-n}\,\times\nn
	&&\qquad \times
	\left[
	\frac{(Z_1\,\partial_{Y_2}+Z_2\,\partial_{Y_1}+Z_1 Z_2\,\partial_G)^{s_{3}-n}}
	{(s_3-n)!}
	\frac{Y_{1}^{s_{1}-n}}{(s_1-n)!}\, \frac{Y_{2}^{s_{2}-n}}{(s_2-n)!}\,
	\frac{G^{n}}{n!}\,
	\right]_{G=G(Y,Z)}\,.\label{Bracket2}
\ea
Since the above expression is homogeneous in the radial coordinate,
one can also replace the powers of $\l$ by
\be
	\l^{s_{3}-n}\quad \to \quad
	(d-5)(d-3)\cdots(d+2\,(s_{3}-n)-7)\, \L^{s_3-n}\,.
	\label{l replace}
\ee
However, in the following we shall keep $\l$ for the brevity of expression, 
unless the above replacement is instructive.

In order to see more transparently the bracket structure,
it is convenient to change our convention.
First, we remove $Z_{1}$ and $Z_{2}$, by 
choosing a convention with the 
evaluation $U_{i}=U$ instead of $U_{i}=0$\,:
\ba
	\big[\hspace{-3pt}\big[\,\bar E_{1}\,,\,
	\bar E_{2}\,
	\big]\hspace{-3pt}\big]_{3}=
	\Pi_{E}\,\mathfrak F^{s_{3}}{}_{s_{1}s_{2}}(Y_{1},Y_{2},Y_{3}\,Z_{3})\, \bar E_{1}\, \bar E_{2}\,
	\big|_{{}^{X_{i}=X}_{U_{i}=U}}\,,
	\label{global}
\ea
where the new bracket function $\mathfrak F^{s_{3}}{}_{s_{1}s_{2}}$ is related to the old one $\cF^{s_{3}}{}_{s_{1}s_{2}}$ by
\be
	\mathfrak F^{s_{3}}{}_{s_{1}s_{2}}(Y_{1},Y_{2},Y_{3}\,Z_{3})
	=\frac1{(1-\partial_{Z_{1}})(1-\partial_{Z_{2}})}\,
	\cF^{s_{3}}{}_{s_{1}s_{2}}(Y_{1},Y_{2},Z_{1},Z_{2},Y_{3}\,Z_{3})\,\Big|_{Z_{1}=0=Z_{2}}\,.	
	\label{remove ZZ}
\ee
Second, using the Killing equation, we remove $Y_{3}$'s at the price of introducing 
$B$-dependence: 
\be
	\mathfrak F^{s_{3}}{}_{s_{1}s_{2}}(Y_{1},Y_{2},B\,Z_{3})
	=e^{\frac12\left[(Y_{1}+Y_{2})\,\partial_{Z_{3}}+
	B\left(\partial_{Y_{1}}+\partial_{Y_{2}}\right)\right]Z_{3}\,\partial_{w}}
	\,\mathfrak F^{s_{3}}{}_{s_{1}s_{2}}(Y_{1},Y_{2},w)\,\big|_{w=0}\,.
\ee
Note that since $Y_{3}$ appears always as $Y_{3}\,Z_{3}$
in the expression \eqref{Bracket2}\,, $B$ also
appears as $B\,Z_{3}$\,.
As we have mentioned before, any cubic interaction leads to consistent (in the sense
of closure) bracket for HS algebra,
and one can show that any consistent bracket should have the form:
\be
	\mathfrak F^{s_{3}}{}_{s_{1}s_{2}}(Y_{1},Y_{2},B\,Z_{3})
	=\mathfrak F^{s_{3}}{}_{s_{1}s_{2}}(G_{3},H_{3})
	\qquad [\,G_3=Y_1+Y_2\,,\ H_3=Y_1\,Y_2+B\,Z_3\,]\,.
	\label{AdS bracket}
\ee
For concreteness sake, we provide several examples of such brackets
obtained from cubic interactions given by coupling constant $k_{n}$'s \eqref{K k}\,
(here we disregard the higher-derivative couplings 
associated with lower $n$'s in \eqref{K k}
which give redundant structures):
\begin{center}
\begin{tabular}{ |c|c|c}
\hline 
  $s_{1}\!-\!s_{2}\!-\! s_{3}$ & Structure constants 
 \\ \hline\hline 
  $2-3-3$ & 
  $\mathfrak F^{3}{}_{23}=k_{2}\,3\,\l\,G_{3}$\,, \quad
  $\mathfrak F^{2}{}_{33}=k_{2}\,G_{3}\,H_{3}$
 \\ \hline
 $3-3-3$ & 
$\mathfrak F^{3}{}_{33}=k_{2}\,\frac12\,\l\,(9\,G_{3}^{2}-4\,H_{3})
+k_{3}\,\frac12\,(3\,G_{3}^{2}-2\,H_{3})$
 \\ \hline
  $3-4-4$ &
   $\mathfrak F^{4}{}_{34}=k_{2}\,3\,\l^{2}\,(3\,G_{3}^{2}-\,H_{3})
   +k_{3}\,3\,\l\,(2\,G_{3}^{2}-H_{3})$\,,
    \\  & 
    $\mathfrak F^{3}{}_{44}=k_{2}\,\frac14\,\l\,
  (9\,G_{3}^{2}\,H_{3}-5\,H_{3}^{2})
  +k_{3}\,\frac12\,(3\,G_{3}^{2}\,H_{3}-2\,H_{3}^{2})$
  \\ \hline
  $4-4-4$ & 
  $\mathfrak F^{4}{}_{44}=k_{3}\,\l\,(8\,G_{3}^{3}-9\,G_{3}\,H_{3})
  +k_{4}\,(2\,G_{3}^{3}-3\,G_{3}\,H_{3})$
  \\ \hline
  $3-4-5$ & $\mathfrak F^{3}{}_{45}=k_3\,\frac12\,G_3\,H_3^2$\,,
  \quad
	$\mathfrak F^{4}{}_{53}=k_3\,4\,\l\,G_3\,H_3$\,,
	\quad
	$\mathfrak F^{5}{}_{34}=k_3\,12\,\l^{2}\,G_3$
  \\ \hline  
\end{tabular}\end{center}
At this stage, let us come back to the issue of classifying
all non-Abelian vertices. 
The point is that the map from  
the space of consistent cubic interactions 
to the space of consistent brackets:
\be
	 \{\,K(Y,G)\,\} \quad \longrightarrow \quad 
	\{\,\mathfrak F^{s_{i}}{}_{s_{i+1}s_{i-1}}(G_{i},H_{i})\,\}\,,
\ee 
is surjective, as one can evince from the above examples. Although we do not provide an explicit demonstration of this point,
it should be clear if one reflects the Fradkin-Vasiliev 
construction in the frame-like approach \cite{Fradkin:1986qy,Fradkin:1987ks,Vasiliev:2011xf,Boulanger:2012dx}.
Hence, one can conclude that, defining $N_{3}$ and $\bar N_{3}$\,
as the numbers of couplings with $[\![ E_{1}, E_{2} ]\!]\neq 0$
and $[\![ \bar E_{1},\bar E_{2} ]\!]\neq 0$ respectively, then
\be
	\bar N_{3} \le N_{3}
	< s_{\rm\sst mid}-\tfrac12\left(s_{\rm\sst max}+s_{\rm\sst mid}-s_{\rm\sst min}\right),
	\label{N<N}
\ee
where the last inequality is obtained counting the number of 
couplings which can be written as $H$-couplings.
Now, let us count the number $\bar N_{3}$
of independent structures in the bracket: 
\be
	\mathfrak F^{s_{3}}{}_{s_{1}s_{2}}(G_{3},H_{3})=
	\sum_{v,h}\,c_{v,h}\,G_{3}^{\,v}\,H_{3}^{\,h}\,.
\ee
Consistency of the number of contractions with the powers of $X$ and $U_{i}$ gives (for more details, see Appendix \ref{Appendix: C})
\be
	v+2\,h=s_{1}+s_{2}-s_{3}-1\,,\qquad
	v+h\le {\rm min}\{s_{1},s_{2}\}-1\,,
\ee
leading to the following bound for $h$\,:
\be
	{\rm max}\{s_{1},s_{2}\}-s_{3}\le h \le \tfrac12\left(s_{1}+s_{2}-s_{3}-1\right),
\ee
so the number of integer $h$'s corresponds to the number of independent structures:
\be
	\bar N_{3}=\left\lceil \tfrac12\,(s_{\rm\sst min}+s_{\rm\sst mid}-s_{\rm\sst max})\right\rceil\,,\label{counting}
\ee
where \mt{\lceil x\rceil={\rm min} \{n\in \mathbb Z\,|\, n\ge x \}}\,.
Finally, from the inequality \eqref{N<N},
one can conclude that $N_{3}=\bar N_{3}$\,.
 
Let us consider the following bilinear form for the Killing tensors:
\be
	\big\langle\,\bar E_{1}\,\big|\,\bar E_{2}\,\big\rangle
	=b_{s_{1}}\,\frac{\left(
	\partial_{U_{1}}\!\cdot \partial_{U_{2}}\right)^{r_{1}}}{r_{1}!}\,
	\frac{\left(\partial_{X_{1}}\!\cdot \partial_{X_{2}}
	\right)^{r_{1}}}{r_{1}!}\,
	\bar E_{1}(X_{1},U_{1})\,\bar E_{2}(X_{2},U_{2})\,
	\Big|_{{}^{X_{i}=0}_{U_{i}=0}}\,,
\ee
given by a series of coefficients $b_{s}$'s. 
This bilinear form become invariant if it makes the structure constant cyclic:
\be
	\big\langle \,\bar E_{3}\,\big|\, \big[\hspace{-3pt}\big[\,\bar E_{1}\,,\,
	\bar E_{2}\,
	\big]\hspace{-3pt}\big]\,\big\rangle
	=\big\langle \,\bar E_{1}\,\big|\, \big[\hspace{-3pt}\big[\,\bar E_{2}\,,\,
	\bar E_{3}\,
	\big]\hspace{-3pt}\big]\,\big\rangle
	=\big\langle \,\bar E_{2}\,\big|\, \big[\hspace{-3pt}\big[\,\bar E_{3}\,,\,
	\bar E_{1}\,
	\big]\hspace{-3pt}\big]\,\big\rangle\,.
	\label{cyclic}
\ee
The bracket structures $\mathcal F^{s_{i}}{}_{s_{i+1}s_{i-1}}$ resulted from cubic interactions satisfy 
this cyclicity condition if $b_{s}$'s satisfy
\be
	\frac{b_{s}}{b_{s-1}}= (d+2s-7)\,\L\,,
\ee
where we have used the replacement \eqref{l replace} to get the above relation.
Note that this condition is independent from the choice of coupling constants $k_{n}$\,, hence all the (A)dS cubic interactions are automatically consistent with 
the condition \eqref{cyclic}.
It is convenient to present the solution of the above equation as
\be
	b_{s}=g^{-1}\,(d-5)(d-3)\cdots (d+2s-7)\,\Lambda^{s}=\frac{\lambda^{s}}g\,.
\ee
Then, one can rescale the fields so that the invariant form become proportional to the identity: \mt{b_{s}=1} --- this would correspond to
the canonical normalization for HS fields. 
As a result, the structure constants become 
automatically cyclic, while the kinetic term takes the form:
\be
	S^{\sst (2)}_{\sst\text{can.}}[\Phi]\ett -\,\frac1{2\,g}\
	\int_{\rm\sst (A)dS}
	e^{\,\l\,\partial_{U_{1}}\!\cdot\,\partial_{U_{2}}}\,
	\Phi(X,U_{1})\,\partial_{X}^{\,2}\,\Phi(X,U_{2})\,
	\Big|_{\sst U_{i}=0}\,,
	\label{amb free canon}
\ee
with an explicit dependence on $\l$ for each different spin.
An important point to remind here is that the rescaling involves the cosmological constant, so that the invariant form is degenerate in the flat-space limit.

The explicit analysis of the brackets provides a one-to-one correspondence between
non-Abelian interacting vertices in the metric-like and in the frame-like formalisms.
The next step towards the systematics of HS interactions requires
the study of the full quartic consistency,
which implies in particular that the consistent brackets 
must satisfy the Jacobi identity.
Hence, one may first  explore all possible consistent HS algebras
(see \cite{Boulanger:2013zza} for the analysis of the algebras involving only symmetric fields,
and \cite{Vasiliev:2012tv,Gelfond:2013xt} for their multi-particle extensions).
To conclude the discussion of (A)dS global symmetries, let us consider the gauge-algebra deformation induced by the cubic interaction associated with the coupling function: 
\be
	K(Y,G)=\frac1{g}\,e^{Y_{1}+Y_{2}+Y_{3}}\,,
\ee
which does not involve any $G$ dependence, and generates only the highest derivative couplings for each \mt{s_{1}\!-\!s_{2}\!-\!s_{3}}.
The corresponding cubic interaction takes the form:
\be
	C=e^{\l\,\cD}\,K
	=\frac1g\,e^{Y_{1}+Y_{2}+Y_{3}+\lambda\,(Z_{1}+Z_{2}+Z_{3})}\,,
	\label{cubic K}
\ee 
which resembles 
the cubic interaction for the excitations of the first Regge trajectory of the open bosonic string (for simplicity we do not explicitly write down Chan-Paton factors):
\be
	C\sim\frac1{g_{o}}\,
	e^{Y_{1}+Y_{2}+Y_{3}+\frac1{\alpha'}\,(Z_{1}+Z_{2}+Z_{3})}\,.
\ee
Interestingly, the gauge-algebra deformation induced by \eqref{cubic K}
gives exactly the Moyal bracket: 
\be
		\big[\hspace{-3pt}\big[\,\bar E^{\,a}_{1}\,,\,
	\bar E^{\,b}_{2}\,
	\big]\hspace{-3pt}\big]^{c}(X,U)={\Pi_{E}}\,
	\left(d^{c}{}_{ab}\,\sinh{G_{3}}+f^{c}{}_{ab}\,\cosh{G_{3}}\right)
	\bar E^{\,a}_{1}(X_{1},U_{1})\,
	\bar E^{\,b}_{2}(X_{2},U_{2})\,\Big|_{{}^{X_{i}=X}_{U_{i}=U}}\,,
\ee
where we have reinstated Chan-Paton factors 
and the corresponding \emph{internal} structure constants --- totally symmetric $d_{abc}$ and totally antisymmetric $f_{abc}$\,.
If the Killing tensors were traceful tensors (so that we do not insert
the projector $\Pi_{E}$),
then the latter would be a good bracket 
that does even satisfy the Jacobi identity.
Notice however that in a setting 
where finitely many irreducible fields are present for given spin (as in the Vasiliev theory),
one should consider only traceless tensors, so that the coupling \eqref{cubic K}
does not actually fulfil the Jacobi identity on traceless tensors.
Nevertheless, it is still challenging to reconsider this type of coupling in a HS theory
with reducible spectrum \cite{Bengtsson:1986ys,Henneaux,Bonelli:2003kh,Sagnotti:2003qa,Francia:2010qp,Campoleoni:2012th,Francia:2013sca}. Actually, such a reducible theory
is a natural candidate for the tensionless limit of the first Regge-trajectory of the open bosonic string.

\subsubsection*{Flat space}

Let us consider now the case of flat-space interactions where the bracket is given by
\be
	\cF^{s_{3}}{}_{s_{1}s_{2}}= \frac14\,
	\left(\partial_{Y_{1}}\partial_{Z_{1}}+\partial_{Y_{2}}\partial_{Z_{2}}\right)
	\sum_{n=0}^{s_{\rm\sst min}}\,k_{n}\,	
	\frac{Y_{1}^{s_{1}-n}}{(s_1-n)!}\, \frac{Y_{2}^{s_{2}-n}}{(s_2-n)!}\,
	\frac{Y_{3}^{s_{3}-n}}{(s_3-n)!}\,\frac{[G(Y,Z)]^{n}}{n!}\,.\label{Bracket2}
\ee
One can again remove $Z_{1}, Z_{2}$ by \eqref{remove ZZ} and
replace the $Y_{3}$-dependence by $B$\,. 
However, at this time,
$Y_{3}$ and $B$ can appear alone  (not always in the combinations $Y_{3}\,Z_{3}$
and $B\,Z_{3}$\,, respectively).
The general form of the consistent bracket structure can be derived also for the flat-space case, and it reads
\be
	\mathfrak F^{s_{3}}{}_{s_{1}s_{2}}(Y_{1},Y_{2},B\,Z_{3},B)
	=\mathfrak F^{s_{3}}{}_{s_{1}s_{2}}(G_{3},H_{3},B)\,.
	\label{flat bracket}
\ee 
One can notice that, compared to the (A)dS case \eqref{AdS bracket},
the flat-space brackets may 
also depend on $B$ in addition to $G_{3}$ and $H_{3}$\,.
Before continuing our discussions,
let us provide some examples for more concrete understanding:
\begin{center}
\begin{tabular}{ |c|c|c}
\hline 
  $s_{1}\!-\!s_{2}\!-\! s_{3}$ & Structure constants 
 \\ \hline\hline 
  $2-3-3$ & 
  $\mathfrak F^{3}{}_{23}=0$\,, \quad
  $\mathfrak F^{2}{}_{33}=k_2\,G_3\,H_3$
 \\ \hline
 $3-3-3$ & 
$\mathfrak F^{3}{}_{33}=-\tfrac32 \, (k_2+a)\,B\,G_3^2+a\,B\,H_3+\tfrac12\,k_3\,(3\,G_3^2-2\,H_3)$
 \\ \hline
  $3-4-4$ &
   $\mathfrak F^{4}{}_{34}=-(k_3+4\,a)\,B\,H_3+6\,a\,B\,G_3^2$\,,
    \\  & 
    $\mathfrak F^{3}{}_{44}=-\tfrac14\,(k_2+16\,b)\,B\,H_3^2+6\,b\,B\,G_3^2\,H_3+\tfrac12\,k_3\,(3\,G_3^2\,H_3-2H_3^2)$
  \\ \hline
  $4-4-4$ & 
  $\mathfrak F^{4}{}_{44}=2\,(k_3+4\,a)\,B\,G_3^3-6\,(k_3+2\,a)\,B\,G_3\,H_3+k_4\,(2\,G_3^3-3\,G_3\,H_3)$
  \\ \hline
  $3-4-5$ & $\mathfrak F^{3}{}_{45}=\tfrac12\,k_3\,G_3\,H_3^2$\,,
  \quad
	$\mathfrak F^{4}{}_{53}=0$\,,
	\quad
	$\mathfrak F^{5}{}_{34}=0$
  \\ \hline  
\end{tabular}\end{center}
As one can see from the appearance of arbitrary 
coefficients $a$ and $b$,
there exist in fact certain functions 
$\mathfrak F^{s_{3}}{}_{s_{1}s_{2}}(G_{3}, H_{3},B)$
which identically vanish when acted on the Killing tensors.
Hence, the description \eqref{flat bracket}
of the flat-space bracket 
may involve some redundancies.
Even after removing such redundancies (by fixing $a$ and $b$), 
there exist in general more bracket structures than non-Abelian couplings.
{To iterate, let us remind the reader that any functions of $G_{3}, H_{3}$ and $B$ provide consistent brackets in the sense that the bracket of two Killing tensors
gives again a Killing tensor. However, not all such brackets can
be associated with consistent cubic interactions: 
look, for instance, the 3\,--\,3\,--\,3 and 4\,--\,4\,--\,4 examples,
which do not involve the structures $G^{2}_{3}$ and $G_{3}^{3}$ respectively.
In other words, not all such brackets can 
be gauged to give consistent cubic interactions in flat space,
as opposed to the (A)dS case.}

Notice that in some cases
such as $\mathfrak F^{3}{}_{23}$, $\mathfrak F^{4}{}_{53}$ 
and $\mathfrak F^{5}{}_{34}$\,, 
there does not exist any non-trivial bracket structure at all. 
This is one of the generic features of flat-space global symmetries
associated with the couplings in Class III in Table~\ref{tab: classification}:
only one among the three structure constants is non-vanishing. 
This implies that the invariant bilinear form is necessarily degenerate if one of 
such couplings is present.  
In this perspective,
of particular interest would be the non-Abelian \mt{s_{1}\!-\!s_{2}\!-\!2} interactions
with \mt{s_{1},s_{2}>2}\,.
They have $s_{1}+s_{2}-2$ derivatives and do fall within Class III. 
Therefore, a consequence of our analysis is that 
HS generators do not rotate under isometry transformations:
\be
	[\![\,T_{s}\,,T_{2}\,]\!] =0 \qquad [\forall s>2]\,,
\ee
where $T_{s}$ stands for a Killing tensor associated with spin-$s$ field.
This implies again that Gravity is not compatible with HS fields 
(recall that we have already shown that HS do not transform accordingly to general covariance).
Despite the lack of general covariance, one can still study
a possible consistency of spin-two non-Abelian (non-gravitational) couplings with HS fields.
Schematically, they induce the following brackets:\footnote{In fact, 
$[\![ T_{s}, T_{2} ]\!]$ can be non-vanishing for $s=3$ 
if the theory involves the minimum derivative $2\!-\!2\!-\!3$ coupling.
In such a case, the analysis of the Jacobi identity becomes more involved 
(see Appendix~\ref{Appendix: D} for the details)
but the conclusions we draw here do not change anyway.}
\be
	[\![\,T_{s}\,,T_{s'}\,]\!] \sim T_{2} + \sum_{s''>2} T_{s''}\,,
	\qquad
	[\![\,T_{s}\,,T_{2}\,]\!] =0\qquad
	[\,\forall\,s,s'>2\,]\,,
	\label{GravAlg}
\ee
where the terms $\sum_{s''>2}T_{s''}$ correspond to possible contributions induced by \mt{s_{1}\!-\!s_{2}\!-\!s_{3}} interactions with all $s_{i}>2$\,. 
Since global symmetries must satisfy the Jacobi identity,
the above leads on the one hand to 
\be
	[\![\,[\![\,T_s\,,\,T_{s'}\,]\!]\,,\,T_2\,]\!]
	=-\,[\![\,[\![\,T_{s'}\,,\,T_2\,]\!]\,,\,T_s\,]\!]
	-[\![\,[\![\,T_2\,,\,T_s\,]\!]\,,\,T_{s'}\,]\!]=0\,,
	\label{Jaco ss2}
\ee
while on the other hand to
\be
	[\![\,[\![\,T_s\,,\,T_{s'}\,]\!]\,,\,T_2\,]\!]
	\sim [\![\,T_2+{\small\sum_{s''>2}}\,T_{s''}\,,\,T_2\,]\!]
	\sim [\![\,T_2\,,\,T_2\,]\!]\,.
\ee
Combining the above two conditions, one arrives to the conclusion: 
\be
	[\![\,T_2\,,\,T_2\,]\!]=0\,.
\ee
Therefore, one can see that the non-Abelian \mt{s_{1}\!-\!s_{2}\!-\!2} 
couplings with $s_{i}>2$ \emph{cannot} pass the Jacobi consistency test in a theory whose spin-two charges form the isometry algebra 
$[\![\,T_2\,,\,T_2\,]\!]\sim T_{2}$\,.\footnote{A similar argument can be applied to certain  $s_{1}\!-\!s_{2}\!-\! s$ non-Abelian interactions: the condition that all non-Abelian coupling
belong to Class III reads $s_{1}+s_{2}\ge 3\,s$\,.
One can also see that when $s=2$\,,
any HS $s_{1}, s_{2} >2$ satisfies this condition.}
In particular, this implies that if a spin-two field couples to HS (even non-gravitationally but through
a non-Abelian coupling), it cannot couple gravitationally to any other field.\footnote{See \cite{Metsaev:1991mt,Metsaev:1991nb} for some examples of amplitudes involving such interactions. 
It would be interesting to clarify these examples 
from the viewpoint of  global-symmetry algebra.}
If one insists on keeping non-Abelian spin-two self-interactions into the game, 
then the only allowed \mt{s_{1}\!-\!s_{2}\!-\!2} interactions with $s_{i}>2$ are the Abelian ones.
In such a theory, non-trivial structure constants may still arise among HS particles,
but the global symmetry becomes just a direct sum between 
the isometry algebra and the algebra generated by HS killing tensors. 
Hence, the latter can be better interpreted as an \emph{internal} symmetry.
This analysis can be viewed as a field-theoretical version of the 
Coleman-Mandula theorem \cite{Coleman:1967ad}.
However, let us emphasize that 
our conclusions rely neither on 
the finiteness of the spectrum below a given mass-scale,
nor on the locality of the Lagrangian.\footnote{The issue of locality may arise only at the quartic order.} 
Let us note also that we assume 
the free action to have the standard (Fronsdal) kinetic term:
diagonal in fields and of two-derivatives. 
This assumption forbids to obtain, for instance, the conformal theory of HS \cite{Segal:2002gd,Bekaert:2010ky} which actually exists in flat-space background ---
the spin $s$ part of its kinetic term has $2\,s$ derivatives.\footnote{Other non-standard higher-derivative kinetic terms for HS fields have been studied in \cite{Joung:2012qy} together with their gauge symmetries.}
It is also worth to comment on the other no-go theorems
based on the S-matrix consistency and/or on the general covariance of HS 
(namely the equivalence principle)
\cite{Weinberg:1964ew,Porrati:2008rm}.
Such results forbid gravitational HS interactions in flat space.
Let us mention however that our analysis does not yet rule out the possibility that 
a non-Abelian self-interacting spin-two interacts with HS fields through Abelian couplings.

\section{Abelian and non-deforming interactions as $H$-couplings} 
\label{sec: GH}

In the previous section, we have seen that while in flat space the question of gauge-symmetry deformation is equivalent to a simple counting of derivatives, in (A)dS 
the same question can be rephrased as the question whether a
cubic interaction $C(Y,Z)$ 
can be expressed making use of  $\tilde K(Y_{\ell}, H_{12}, H_{23}, H_{31})$
or not. More precisely, depending 
on the possible values of $\ell$\,, such coupling leads to
\begin{itemize}
\item trivial deformation of the $i$-th gauge transformation:
\mt{\delta^{\sst (1)}_{\sst E_{i}}=0}, if and only if
the re-expression is possible for $\ell=i+1$ or $i-1$\,;
\item trivial deformation of gauge algebra: 
\mt{[\![ E_{i},E_{j} ]\!]^{\sst(0)}=0},
if and only if the re-expression is possible for any $\ell=1,2,3$.
\end{itemize}
In this section, we investigate these conditions and show, in particular, how
they can be interpreted as conditions on the maximum number
of derivatives in the (A)dS couplings.

\subsection{Contraction of curls}

Before analyzing the couplings given by the operators $H_{ij}$'s,
let us make a few remarks on the physical meaning of the $H_{ij}$'s.
From their definitions \eqref{H}, it is easy to notice that 
they can be built using matrix operators 
$[\partial_{X_{i}}\partial_{U_{i}}]$ with components:
\be
	[\partial_{X_{i}}\partial_{U_{i}}]_{\sst MN}
	=\partial_{X_{i}^{\sst [M}}\partial_{U_{i}^{\sst N]}}\,.
	\label{curl}
\ee
These are nothing but the curl operators with respect to the $i$-the field, and they are manifestly gauge invariant.
One can now consider gauge-invariant scalar operators by
taking traces of such operators (that is, contracting different curls) as
\be
	H_{i_{1}\cdots i_{n}}={\rm Tr}\big(\,[\partial_{X_{i_{1}}}\partial_{U_{i_{1}}}]
	\cdots
	[\partial_{X_{i_{n}}}\partial_{U_{i_{n}}}]\, \big)\,.
	\label{Hs}
\ee
However, as \mt{[\partial_{X}\partial_{U}]^{\otimes n}}
corresponds to the curvature operator of rank $2n$\,,
one can check that
the only independent ones among $H_{i_{1}\cdots i_{n}}$'s are 
\be
	H_{12}\,,\quad H_{23}\,,\quad H_{31}\,,\quad H_{123}\,.
\ee
Moreover, the square of $H_{123}$ reduces to 
a product of $H_{ij}$'s as
\be
	H_{123}^{\,2}\,\Phi_{123}\overset{\rm\sst TT}{\approx}\tfrac18\,H_{12}\,H_{23}\,H_{31}\,\Phi_{123}\,,
\ee
and $H_{123}$ itself can be expressed in terms of $H_{ij}$ as
\be
	H_{123}\,\Phi_{123}\overset{\rm\sst TT}{\approx}
	\tfrac12\,
	Y_{i}\,H_{i+1\,i+2}\,\Phi_{123}\,,
	\label{H123}
\ee
when acted on the three-local fields $\Phi_{123}(X_{1},U_{1}; X_{2},U_{2}; X_{3},U_{3})$ satisfying
the homogeneity and tangentiality conditions on each of the points.

\subsection{$H$-couplings}
\label{sec: H}

In this subsection, we closely look the $H$-couplings
$\tilde K(Y_{\ell}, H_{12}, H_{23}, H_{31})$\,,
and show how many derivatives they contain.
For that, we need to find an independent set
of couplings which span the space of all $H$-couplings.
Such a set can be obtained by considering the monomials (see Appendix~C of \cite{Joung:2012hz} for more details):
\ba
	&& Y_{\ell}^{\,\s}\,\tilde H_{1}^{\,h_{1}}\,\tilde H_{2}^{\,h_{2}}\,
	\tilde H_{3}^{\,h_{3}} \nn
	&& \overset{\sst\rm TT}{\approx}
	Y_{\ell}^{\,\s} \prod_{i=1}^{3}
	\left[
	\sum_{n_{i}=0}^{h_{i}} \binom{h_{i}}{n_{i}}
	\big[h_{i}-\epsilon_{\ell,i}\,\s\big]_{n_{i}}
	\big(\l\,Z_{i}\big)^{n_{i}}
	\big(Y_{i+1}\,Y_{i-1}\big)^{h_{i}-n_{i}}\right],
	\label{YHHH}
\ea
where $\e_{\ell,i}=2\,\delta_{\ell,i}-1$ and $\tilde H_{i}$'s are
\be
	\tilde H_{1}:=-H_{23}\,,\qquad
	\tilde H_{2}:=-H_{31}\,,\qquad
	\tilde H_{3}:=-H_{12}\,.
\ee
In eq.~\eqref{YHHH}, we have also provided their expressions as functions of $Y$ and $Z$\,,
making clear the counting of derivatives (the $Y_{i}$'s contain one derivative while the $Z_{i}$'s do not). 
The powers $\s, h_{1}, h_{2}, h_{3}$ of \eqref{YHHH}
are related to the spins by
\be
	s_{i}=h_{i+1}+h_{i-1}+\delta_{i,\ell}\,\s\,.
	\label{spins}
\ee
For the following analysis, it is also convenient to distinguish
two cases --- whether the spins satisfy a triangular inequality or not:
\be
	s_{\rm\sst max}\le s_{\rm\sst mid}+s_{\rm\sst min}
	\qquad {\rm or} \qquad
	s_{\rm\sst max}> s_{\rm\sst mid}+s_{\rm\sst min}\,.
	\label{Triang}
\ee
However, it will turn out that the second case can be considered as a special 
class of the first one.

\subsubsection*{Triangular case: 
$s_{\rm\sst max}\le s_{\rm\sst mid}+s_{\rm\sst min}$}

When the triangular inequality holds, from eq.~\eqref{spins}, 
we can determine the $h_{i}$'s in terms of the spins and $\s=\u+2n$\,,
where $\u:={\rm mod}_{2}(s_{1}+s_{2}+s_{3})$
and the non-negative integer $n$ is bounded as
\be
	n\le n_{*}:={\rm min}\left\{\tfrac{s_{\ell}+s_{\ell+1}-s_{\ell-1}}2\,,
	\tfrac{s_{\ell}+s_{\ell-1}-s_{\ell+1}}2\right\}.
	\label{n*}
\ee
Hence, we obtain a class of couplings labelled by $n$\,:\footnote{The 
\mt{n=0} case is special since 
we can apply the identity \eqref{H123}
to replace $Y_{\ell}\,H_{\ell+1\,\ell-1}$ with $H_{123}$\,.
Involving only $H_{ij}$'s and $H_{123}$\,, 
this coupling correspond to the full contraction of maximum curls,
namely three linearized curvatures.
So, it is the Born-Infeld coupling.}
\be
	(Y_{\ell}\,\tilde H_{\ell})^{\u}\,Y_{\ell}^{2n}\,
	\tilde H_{1}^{\lfloor\frac{s_{2}+s_{3}-s_{1}}2\rfloor+\e_{\ell,1}\,n}\,
	\tilde H_{2}^{\lfloor\frac{s_{3}+s_{1}-s_{2}}2\rfloor+\e_{\ell,2}\,n}\,
	\tilde H_{3}^{\lfloor\frac{s_{1}+s_{2}-s_{3}}2\rfloor+\e_{\ell,3}\,n}\,,
	\label{Hn}
\ee
where \mt{\lfloor x\rfloor:={\rm max} \{m\in \mathbb Z\,|\, m \le x \}}\,.
The highest-derivative parts of these couplings are all the same:
$Y_{1}^{s_{1}}\,Y_{2}^{s_{2}}\,Y_{3}^{s_{3}}$\,,
and its number of derivatives is $s_{1}+s_{2}+s_{3}$ --- the maximum possible value. By taking linear combinations of these $n_{*}+1$ couplings,
we can make $n_{*}$ different couplings where the highest-derivative terms cancel. The new highest-derivative part of such couplings
are again all the same: $Y_{1}^{s_{1}-1}\,Y_{2}^{s_{2}-1}\,Y_{3}^{s_{3}-1}\,G(Y,Z)$\,, and their number of derivatives is lowered by two: $s_{1}+s_{2}+s_{3}-2$\,.
One can continue this procedure until to obtain
a coupling whose highest-derivative part 
is $Y_{1}^{s_{1}-n_{*}}\,Y_{2}^{s_{2}-n_{*}}\,Y_{3}^{s_{3}-n_{*}}\,G(Y,Z)^{n_{*}}$
(with the number of derivative $s_{1}+s_{2}+s_{3}-2\,n_{*}$).
Hence, starting from the basis \eqref{Hn} of $H$-couplings,
we can obtain a new basis where couplings have all different
highest-derivative parts. The couplings 
whose number of highest derivatives is 
\mt{s_{1}+s_{2}+s_{3}-2\,n} reads
\ba
	Q^{\sst [n]}_{\ell,s_{1}s_{2}s_{3}} \eq
	q^{\sst [n]}_{\ell,\u}\,
	(Y_{\ell}\,\tilde H_{\ell})^{\u}\,
	\tilde H_{1}^{\lfloor\frac{s_{2}+s_{3}-s_{1}}2\rfloor-\bar\d_{\ell,1}\,n}\,
	\tilde H_{2}^{\lfloor\frac{s_{3}+s_{1}-s_{2}}2\rfloor-\bar\d_{\ell,2}\,n}\,
	\tilde H_{3}^{\lfloor\frac{s_{1}+s_{2}-s_{3}}2\rfloor-\bar\d_{\ell,3}\,n}\,,
	\nn
	&\!\overset{\rm\sst TT}{\approx}\!&
	Y_{1}^{s_{1}-n}\,Y_{2}^{s_{2}-n}\,Y_{3}^{s_{3}-n}\,G(Y,Z)^{n}+
	\cO\big(\l)\,,
\ea
where $\bar\d_{\ell,i}=1-\delta_{\ell,i}$ and $q^{\sst [n]}_{\ell,\s}$ is given by
\ba
	&& q^{\sst [n+1]}_{\ell,\s}=
	\big(\tilde H_{\ell+1}\,\tilde H_{\ell-1}-Y_{\ell}^{2}\,\tilde H_{\ell}\big)
	\times \nn
	&&\quad \times\,
	\sum_{k=0}^{n} \binom{n}{k} 
	\frac{(\s+2n+1)_{k}\,[\s+1]_{n-k}}
	{\l^{n+1}\,(\s+1)_{2n+1}}
	\,(\tilde H_{\ell+1}\,\tilde H_{\ell-1})^{k}\,(-Y_{\ell}^{2}\,\tilde H_{\ell})^{n-k}\,.
\ea
To iterate, any $H$-coupling $\tilde K(Y_{\ell}, H_{12}, H_{23}, H_{31})$
can be decomposed in terms of $Q_{\ell}^{\sst [n]}$'s\footnote{Herefrom,
we drop the subscript $s_{1}s_{2}s_{3}$ for compactness sake.}
with $n=0,1,\ldots,n_{*}$\,,
and the value $n_{*}$ depends on $\ell$ \eqref{n*}. 
To be more concrete, let us assume 
without loss of generality
\be
	s_{1}\ge s_{2} \ge s_{3}\,.
\ee
Then, all possible $H$-couplings are given by
\ba
	Q^{\sst [n]}_{1} \quad \left[n\le \tfrac{s_{1}+s_{3}-s_{2}}2\right]; 
	\qquad
	Q^{\sst [n]}_{2}\,,\, Q^{\sst [n]}_{3}
	\quad \left[n\le \tfrac{s_{2}+s_{3}-s_{1}}2 \right]. 
	\label{K1}
\ea
From the above, we can conclude the following:
\begin{itemize}
\item {\bf\underline{Non-deforming couplings}} \\[3pt] First, in the range
\be
	n \le \tfrac{s_{2}+s_{3}-s_{1}}2
	\quad \Rightarrow \quad
	\#_{\partial}\ge 2\,s_{1}\,,
\ee
all three $Q_{1}^{\sst [n]}, Q_{2}^{\sst [n]}$ and $Q_{3}^{\sst [n]}$
are available. In fact, $Q_{2}^{\sst [n]}\overset{\rm\sst TT}\approx Q_{3}^{\sst [n]}$
and they can be expressed as linear combinations of $Q_{1}^{\sst [n]}$'s.\footnote{For
the proof, consider the coupling:
\be
	C=Q_{2}^{\sst [n]}-Q_{1}^{\sst [n]}-c\,Q_{1}^{\sst [n+1]}-
  	\cdots-c'\,Q_{1}^{[{\sst \lfloor (s_{1}+s_{3}-s_{2})/2 \rfloor}]}\,,
	\label{Q2 Q1}
\ee
where the coefficients $c,\ldots, c'$ are chosen such that 
the number of derivatives in $C$ is the lowest possible.
Since $C$ gives $\delta^{\sst (1)}_{\sst E_{3}}=0$\,, it can be 
expressed either by $Q_{1}^{\sst [m]}$'s or by $Q_{2}^{\sst [m]}$'s,
but the number of derivatives of $C$ is too low to allow them.
Hence, one can conclude that $C$ must be zero, that is, 
any $Q_{2}^{\sst [n]}$ can be written
as $Q_{1}^{\sst [m]}$'s. One can do the same for $Q_{3}^{\sst [n]}$
to conclude the same.} 
Hence, the couplings given by $Q_{2}^{\sst [n]}$'s or $Q_{3}^{\sst [n]}$'s 
leave all three gauge transformations undeformed:
 \mt{\delta^{\sst (1)}_{\sst E_{1,2,3}}=0}\,.

\item  {\bf \underline{Deforming Abelian couplings}} \\[3pt]
In the range:
\be
	\tfrac{s_{2}+s_{3}-s_{1}}2<n\le \tfrac{s_{1}+s_{3}-s_{2}}2
	\quad \Rightarrow \quad
	2\,s_{2}\le \#_{\partial}<2\,s_{1}\,,
	\label{middle n}
\ee
only $Q^{\sst [n]}_{1}$ is available but neither $Q^{\sst [n]}_{2}$
nor $Q^{\sst [n]}_{3}$\,.
Hence, the couplings given by $Q^{\sst [n]}_{1}$'s
with $n$ in \eqref{middle n}
necessarily deform  the gauge transformation $\delta^{\sst (1)}_{\sst E_{1}}\neq 0$\,, but leave the other transformations and gauge algebra
undeformed:
$\delta^{\sst (1)}_{\sst E_{2,3}}=0$ and $[\![ E_{i},E_{j} ]\!]^{\sst (0)}=0$\, for all $i$ and $j$.

\item  {\bf \underline{Non-Abelian couplings}} \\[3pt]
Finally, in the range:
\be
	n> \tfrac{s_{1}+s_{3}-s_{2}}2
	\quad\Rightarrow \quad
	\#_{\partial}<2\,s_{2}\,,
\ee
all three $Q_{1}^{\sst [n]}, Q_{2}^{\sst [n]}$ and $Q_{3}^{\sst [n]}$
are not available. 
Hence, any coupling, whose number of highest derivatives 
is smaller than $2\,s_{2}$\,, cannot be written as a $H$-coupling,
and deforms in (A)dS all three gauge transformations
and all gauge algebra commutators:
$\delta^{\sst (1)}_{\sst E_{1,2,3}}\neq0$
and $[\![ E_{i},E_{j} ]\!]^{\sst (0)}\neq 0$\,.
\end{itemize}
This result is summarized in Table \ref{tab: classification},
and one can also count the numbers of couplings
in the above three different classes: 
\ba
	&& \big(\,\textrm{$\#$ of  non-deforming couplings}\,\big)
	=\left\lfloor\tfrac12\,(s_{\rm\sst min}+s_{\rm\sst mid}-s_{\rm\sst max})
	\right\rfloor+1\,,\nn
	&& \big(\,\textrm{$\#$ of  deforming Abelian couplings}\,\big)
	=s_{\rm\sst max}-s_{\rm\sst mid}\,,\nn
	&& \big(\,\textrm{$\#$ of  non-Abelian couplings}\,\big)
	=\left\lceil \tfrac12\,(s_{\rm\sst min}+s_{\rm\sst mid}-s_{\rm\sst max})
	\right\rceil,
\ea
where 
\mt{\lfloor x\rfloor={\rm max} \{m\in \mathbb Z\,|\, m \le x \}}
and \mt{\lceil x\rceil={\rm min} \{n\in \mathbb Z\,|\, n\ge x \}}\,.

\subsubsection*{Non-triangular case: 
$s_{\rm\sst max}> s_{\rm\sst mid}+s_{\rm\sst min}$}
Let us again assume $s_{1}\ge s_{2}\ge s_{3}$\,.
In this case, eq.~\eqref{spins} admit solutions only when $\ell=1$\,,
and the corresponding $H$-couplings can be 
decomposed by
\be
	Q_{1,s_{1}s_{2}s_{3}}^{\sst [n]}
	=q^{\sst [n]}_{1,s_{1}-s_{2}-s_{3}}\,Y_{1}^{s_{1}-s_{2}-s_{3}}\,
	\tilde H_{3}^{\,s_{2}-n}\,
	\tilde H_{2}^{\,s_{3}-n}\,,
	\label{Hn}
\ee
where \mt{n=0,1, \ldots, s_{3}}.
Hence, in the non-triangular case, all the consistent couplings
can be expressed in terms of $Q_{1}^{\sst [n]}$'s
so that they do not deform  $\delta^{\sst (1)}_{\sst E_{2,3}}$
but deform $\delta^{\sst (1)}_{\sst E_{1}}$\,.
This result is summarized in Table \ref{tab: non-triang}.
\begin{table}[hhh]\centering
\begin{tabular}{ |c| c || c | c | c || c|}
  \hline                    
  $n$ & $\#_{\partial}$ & $\delta^{\sst (1)}_{\sst E_{1}}$ & $\delta^{\sst (1)}_{\sst E_{2}}$ 
  & $\delta^{\sst (1)}_{\sst E_{3}}$ & 
  $C^{\sst (3)}$\\[2pt]
  \hline
0  &  $s_{1}+s_{2}+s_{3}$ & \cellcolor{red!25}$\neq0$ & \cellcolor{blue!25}
$=0$ & \cellcolor{blue!25}$=0$ &   \\ 
\vdots  &  $\vdots$ & \cellcolor{red!25}$\vdots$ & \cellcolor{blue!25}\vdots & \cellcolor{blue!25}\vdots & $\approx 
\tilde K(Y_{1},H_{12}, H_{23}, H_{31})$ \\ 
$s_{3}$  &   $s_{1}+s_{2}-s_{3}$ &  \cellcolor{red!25}\vdots &  
\cellcolor{blue!25} \vdots 
& \cellcolor{blue!25}\vdots &  \\
\hline    
\end{tabular}
\caption{Classification of cubic interactions 
(\mt{s_{1}\ge s_{2} \ge s_{3}})
in the non-triangular case (\mt{s_{1}>s_{2}+s_{3}}) 
according to the deformations of gauge transformations and gauge algebras.}
\label{tab: non-triang}
\end{table}

In this section, we have shown how the $H$-couplings are
related to the curl of fields, and that 
in order to reproduce all couplings in Class I and II,
we need to take particular combinations of such functions
where the highest-derivative terms cancel.
From the former point, one can notice that such couplings admit
expressions in terms of HS curvatures. 
Let us notice that the construction of this kind of couplings 
making use of curvatures and of the cancellation of the highest-derivative piece
were also discussed in \cite{Vasiliev:2011xf}.

\subsection{Examples}

Finally, for concreteness sake, let us consider two examples: 
\mt{3\!-\!2\!-\!2} and \mt{3\!-\!3\!-\!2} interactions.  
The Table \ref{tab: classification} reduces then to
Table \ref{tab: 322} and \ref{tab: 332}\,.
\begin{table}[h]
\centering
\parbox{.45\linewidth}{
\centering
\begin{tabular}{ |c| c || c | c | c | }
  \hline                    
  $n$ & $\#_{\partial}$ & $\delta^{\sst (1)}_{\sst E_{1}}$ & $\delta^{\sst (1)}_{\sst E_{2}}$ 
  & $\delta^{\sst (1)}_{\sst E_{3}}$  \\[2pt]
  \hline
0  &  7 & \cellcolor{blue!25}0 & \cellcolor{blue!25}0
& \cellcolor{blue!25}0  \\ 
1  &  5 & \cellcolor{red!25}$*$ & \cellcolor{blue!25}0 & \cellcolor{blue!25}0 \\ 
2  &   3 &  \cellcolor{red!25}$*$ & \cellcolor{red!25}$*$ & \cellcolor{red!25}$*$  \\
\hline    
\end{tabular}
\caption{ $3\!-\!2\!-\!2$ interactions}
\label{tab: 322}
}
\parbox{.45\linewidth}{
\centering
\begin{tabular}{ |c| c || c | c | c |}
  \hline                    
  $n$ & $\#_{\partial}$ & $\delta^{\sst (1)}_{\sst E_{1}}$ & $\delta^{\sst (1)}_{\sst E_{2}}$ 
  & $\delta^{\sst (1)}_{\sst E_{3}}$  \\[2pt]
  \hline
0  &  8 & \cellcolor{blue!25}0 & \cellcolor{blue!25}0
& \cellcolor{blue!25}0  \\ 
1  &  6 & \cellcolor{blue!25}0 & \cellcolor{blue!25}0 & \cellcolor{blue!25}0 \\ 
2  &   4 &  \cellcolor{red!25}$*$ & \cellcolor{red!25}$*$ & \cellcolor{yellow!30}$\L$ \\
\hline    
\end{tabular}
\caption{ $3\!-\!3\!-\!2$ interactions}
\label{tab: 332}}
\end{table}

\subsubsection*{$\bm{3\!-\!2\!-\!2}$ interactions}

Any gauge invariant cubic vertices of $3\!-\!2\!-\!2$ interactions 
can be expanded in the basis of $P^{\sst [n]}_{322}$'s \eqref{Cn} as
\be
	C(Y,Z)=k_{0}\,P^{\sst [0]}
	+k_{1}\,P^{\sst [1]}+k_{2}\,P^{\sst [2]}\,,
	\label{322 exp}
\ee
where we have omitted the subscript $322$\,, and 
the $P^{\sst [n]}$'s are given by  
{\footnotesize
\ba
	P^{\sst [0]} \eq Y_{1}^{\,3}\,Y_{2}^{\,2}\,Y_{3}^{\,2}
	+2\,\l\,Y_{1}^{\,2}\,Y_{2}\,Y_{3}\,
	(2\,Y_{1}\,Z_{1}+3\,Y_{2}\,Z_{2}+3\,Y_{3}\,Z_{3})\nn
	&&+\,2\,\l^{2}\,Y_{1}
	\left[ (Y_{1}\,Z_{1})^{2}+3\,(Y_{2}\,Z_{2})^{2}
	+3\,(Y_{3}\,Z_{3})^{2}
	+6\,Y_{1}\,Z_{1}\,(Y_{2}\,Z_{2}+Y_{3}\,Z_{3})
	+12\,Y_{2}\,Z_{2}\,Y_{3}\,Z_{3}
	\right]
	\nn
	&&+\,12\,\l^{3}
	\left(2\,Y_{1}\,Z_{1}+Y_{2}\,Z_{2}+Y_{3}\,Z_{3}\right)
	Z_{2}\,Z_{3}\,,\nn
	P^{\sst [1]} \eq Y_{1}^{\,2}\,Y_{2}\,Y_{3}\,(Y_{1}\,Z_{1}
	+Y_{2}\,Z_{2}+Y_{3}\,Z_{3}) \nn
	&&+\,\l\,Y_{1}
	\left[ (Y_{1}\,Z_{1})^{2}+2\,(Y_{2}\,Z_{2})^{2}
	+2\,(Y_{3}\,Z_{3})^{2}
	+4\,Y_{1}\,Z_{1}\,(Y_{2}\,Z_{2}+Y_{3}\,Z_{3})
	+6\,Y_{2}\,Z_{2}\,Y_{3}\,Z_{3}
	\right]
	\nn
	&&+\,4\,\l^{2}
	\left(2\,Y_{1}\,Z_{1}+Y_{2}\,Z_{2}+Y_{3}\,Z_{3}\right)Z_{2}\,Z_{3}\,,
	\nn
	P^{\sst [2]} \eq Y_{1}\,(Y_{1}\,Z_{1}+Y_{2}\,Z_{2}+Y_{3}\,Z_{3})^{2} 
	+2\,\l\left(2\,Y_{1}\,Z_{1}+Y_{2}\,Z_{2}+Y_{3}\,Z_{3}\right)
	Z_{2}\,Z_{3}\,.
	\label{P322}
\ea}
\!On the other hand, $H$-couplings provide
a non-deforming and a deforming Abelian vertices,
$Q_{\ell}^{\sst [0]}$ and $Q_{1}^{\sst [1]}$\,.
Being gauge invariant, they can be also expressed as the expansion \eqref{322 exp}:
\ba
	&& Q^{\sst [0]}_{\ell}
	=Y_{\ell}\,\tilde H_{\ell}\,\tilde H_{2}\,\tilde H_{3}
	\approx
	P^{\sst [0]}-4\,\l\,P^{\sst [1]}
	+2\,\l^{2}\,P^{\sst [2]}\,,\nn
	&& Q^{\sst [1]}_{1}=q^{\sst [1]}_{1,1}\,
	Y_{1}\,\tilde H_{1}
	=\l^{-1}\left(Y_{1}^{\,3}\,\tilde H_{1}^{\,2}-
	Y_{1}\,\tilde H_{1}\,\tilde H_{2}\,\tilde H_{3}\right)
	\approx
	P^{\sst [1]}-2\,\l\,P^{\sst [2]}\,.
\ea
These correspond to two vectors in the linear span of \eqref{P322},
and the remaining independent vector $P^{\sst [2]}$ 
represents the non-Abelian coupling of $3\!-\!2\!-\!2$ interactions.

\subsubsection*{$\bm{3\!-\!3\!-\!2}$ interactions}

In the case of $3\!-\!3\!-\!2$ interactions, we have 
{\footnotesize
\ba
	P^{\sst [0]} \eq Y_{1}^{\,3}\,Y_{2}^{\,3}\,Y_{3}^{\,2}
	+3\,\l\,Y_{1}^{\,2}\,Y_{2}^{\,2}\,Y_{3}\,
	(2\,Y_{1}\,Z_{1}+2\,Y_{2}\,Z_{2}+3\,Y_{3}\,Z_{3}) \nn
	&&
	+\,6\,\l^{2}\,Y_{1}\,Y_{2}
	\left[ (Y_{1}\,Z_{1})^{2}+(Y_{2}\,Z_{2})^{2}+3\,(Y_{3}\,Z_{3})^{2}
	+3\,Y_{1}\,Z_{1}\,Y_{2}\,Z_{2}
	+6\,(Y_{1}\,Z_{1}+Y_{2}\,Z_{2})\,Y_{3}\,Z_{3}
	\right]
	\nn
	&&+\,6\,\l^{3}\,Z_{3}
	\left[ 3\,(Y_{1}\,Z_{1})^{2}+3\,(Y_{2}\,Z_{2})^{2}
	+(Y_{3}\,Z_{3})^{2}
	+12\,Y_{1}\,Z_{1}\,Y_{2}\,Z_{2}
	+6\,(Y_{1}\,Z_{1}+Y_{2}\,Z_{2})\,Y_{3}\,Z_{3}\right]
	\nn
	&&+\,36\,\l^{4}\,Z_{1}\,Z_{2}\,Z_{3}^{\,2}\,,
	\nn
	P^{\sst [1]} \eq Y_{1}^{\,2}\,Y_{2}^{\,2}\,Y_{3}\,
	(Y_{1}\,Z_{1}+Y_{2}\,Z_{2}+Y_{3}\,Z_{3})\nn
	&&+\,\l\,Y_{1}\,Y_{2}
	\left[ 2\,(Y_{1}\,Z_{1})^{2}+2\,(Y_{2}\,Z_{2})^{2}
	+4\,(Y_{3}\,Z_{3})^{2}
	+5\,Y_{1}\,Z_{1}\,Y_{2}\,Z_{2}
	+8\,(Y_{1}\,Z_{1}+Y_{2}\,Z_{2})\,Y_{3}\,Z_{3}\right]
	\nn
	&&+\,2\,\l^{2}\,Z_{3}
	\left[ 3\,(Y_{1}\,Z_{1})^{2}+3\,(Y_{2}\,Z_{2})^{2}
	+(Y_{3}\,Z_{3})^{2}
	+10\,Y_{1}\,Z_{1}\,Y_{2}\,Z_{2}
	+5\,(Y_{1}\,Z_{1}+Y_{2}\,Z_{2})\,Y_{3}\,Z_{3}\right]
	\nn
	&&
	+\,10\,\l^{3}\,Z_{1}\,Z_{2}\,Z_{3}^{\,2}\,,
	\nn
	P^{\sst [2]} \eq Y_{1}\,Y_{2}\,(Y_{1}\,Z_{1}+Y_{2}\,Z_{2}+Y_{3}\,Z_{3})^{2} 
	\nn
	&&+\,\l\,Z_{3}
	\left[ 3\,(Y_{1}\,Z_{1})^{2}+3\,(Y_{2}\,Z_{2})^{2}
	+(Y_{3}\,Z_{3})^{2}
	+8\,Y_{1}\,Z_{1}\,Y_{2}\,Z_{2}
	+4\,(Y_{1}\,Z_{1}+Y_{2}\,Z_{2})\,Y_{3}\,Z_{3}\right]
	\nn
	&&
	+\,4\,\l^{2}\,Z_{1}\,Z_{2}\,Z_{3}^{\,2}\,.
\ea}
\!\!On the other hand, $H$-couplings provide two non-deforming vertices,
$Q_{\ell}^{\sst [0]}$ and $Q_{\ell}^{\sst [1]}$\,,
which can be expressed as 
\ba
	&& Q^{\sst [0]}_{\ell}=
	\tilde H_{1}\,\tilde H_{2}\,\tilde H_{3}^{\,2}
	\approx 
	P^{\sst [0]}-5\,\l\,P^{\sst [1]}
	+4\,\l^{2}\,P^{\sst [2]}\,,\nn
	&& Q^{\sst [1]}_{\ell}=q^{\sst [1]}_{\ell,0}\,\tilde H_{\ell}\,\tilde H_{3}
	=\l^{-1}\left(Y_{\ell}^{\,2}\,\tilde H_{\ell}^{\,2}\,\tilde H_{3}-
	\tilde H_{1}\,\tilde H_{2}\,\tilde H_{3}^{\,2}\right)
	\approx
	P^{\sst [1]}-2\,\l\,P^{\sst [2]}\,.
\ea
Again, the other independent coupling $P^{\sst [2]}$ represents
the non-Abelian $3\!-\!3\!-\!2$ interaction.

\section{Discussions}
\label{sec: Conclusions}

In the present paper, we have analysed the deformations of both gauge transformations and gauge algebras induced by the cubic couplings of massless fields constructed in \cite{Joung:2011ww}. The main results of the paper 
are already summarized by Table \ref{tab: classification}, but
let us rephrase them once again here:
\begin{itemize}
\item
Defining 
\mt{\S_{i}=s_{i}} in flat space
and \mt{\S_{i}={\rm max}\{s_{i},s_{\rm \sst mid}\}} in (A)dS, 
the highest number derivatives $\#_{\partial}$
of a coupling with $\delta^{\sst (1)}_{\sst E_{i}}=0$ satisfies
\be
	\big(\,\textrm{$\#_{\partial}$ of a 
	coupling with $\delta^{\sst (1)}_{\sst E_{i}}=0$}\,\big) \ge 
	2\,\S_{i}\,.
\ee
Equivalently, if the highest-derivative term of a  cubic interaction
involves less than $2\,\S_{i}$ derivatives, then the coupling necessarily induces a non-trivial deformation of
gauge transformation: $\delta^{\sst (1)}_{\sst E_{i}}\neq 0$\,.
\item
Defining 
\mt{\S_{ij}={\rm min}\{s_{i},s_{j}\}} in flat space
and \mt{\S_{ij}=s_{\rm \sst mid}} in (A)dS, 
the highest number of derivatives $\#_{\partial}$ in a coupling with
$[\![E_{i},E_{j}]\!]^{\sst (0)}= 0$ satisfies
\be
	\big(\,\textrm{$\#_{\partial}$ of a coupling 
	with }[\![E_{i},E_{j}]\!]^{\sst (0)}= 0\,\big) 
	\ge 2\,\S_{ij}\,.
\ee
Equivalently, if the highest-derivative terms of 
a cubic interaction involves less than $2\,\S_{ij}$ derivatives,
then the coupling necessarily induces a non-Abelian deformation of the
gauge algebra: $[\![E_{i},E_{j}]\!]^{\sst (0)}\neq 0$\,.

\end{itemize}
This is the first classification of cubic interactions
according to the deformations of each bracket $[\![E_{i},E_{j}]\!]^{\sst (0)}$
and each gauge transformation $\delta^{\sst (1)}_{\sst E_{i}}$\,.
As a result, we categorized all cubic interactions into four different classes (see Table \ref{tab: classification}).\footnote{{At this point, let us summarize
once again what was known before about 
the general classification of the HS gauge-symmetry deformations. 
The results can be categorized into two groups --- the metric-like formulation for flat space interactions,
and the frame-like formulation for strictly (A)dS interactions.
In the former case, the minimum number of derivatives 
of cubic interactions with 
$[\![E_{1},E_{2}]\!]^{\sst (0)}=[\![E_{2},E_{3}]\!]^{\sst (0)}=[\![E_{3},E_{1}]\!]^{\sst (0)}=0$ has been identified in \cite{Boulanger:2008tg},
but neither for each of the $[\![E_{i},E_{j}]\!]^{\sst (0)}$'s nor 
for the $\delta^{\sst (1)}_{\sst E_{i}}$'s\,.
In the latter case, the non-Abelian vertices, corresponding in our classification to the Class III and IV couplings,
have been constructed in \cite{Vasilev:2011xf} 
together with 
the discussions on the Abelian and the Current vertices,
which correspond respectively to the Class I and II couplings.}} 
Class I and IV correspond to 
the set of couplings which deform nothing and everything, respectively.
The other two classes have some interesting features.
Class II couplings leave the gauge algebra invariant
but deform one out of three gauge transformations.
This means that the other two fields are charged with respect to the 
last field, and should correspond the couplings responsible for 
HS-algebra multiplets, together with the massless-massive-massive cubic interactions (see below).
Class III couplings induce non-trivial deformations of gauge transformations 
for all three fields (hence for all three brackets) in (A)dS, but
one of them (two of them, considering brackets) vanishes
in the flat-space limit.
The gravitational interactions of HS fields belong to this class,
and the corresponding gauge transformations,
which vanish in the flat-space limit,
are the general coordinate transformation (or diffeomorphism) of HS fields.
The couplings belonging to Class III have also interesting implications on the global symmetries: 
a conclusion similar to the Coleman-Mandula theorem
can be derived in our setting as a consequence of the Jacobi identity. 

\subsection*{$H$- and $G$-couplings}

\begin{figure}[h]
\centering
\begin{tikzpicture}
\draw[red,fill=red!15,thick] (0,-0.3)  ellipse (1.5cm and 1cm);
\node at (0,-0.8) {$G$-couplings};
\draw[blue,fill=blue!15,thick] (0,0)  ellipse (1.2cm and 0.5cm);
\node at (0,0) {$H$-couplings};
\node at (0, 1.2)  {$\mu_{2}-\mu_{3}=0$};

\draw[red,fill=red!15,thick] (7,-0.7)  ellipse (1.4cm and 0.7cm);
\node at (7.1,-0.9) {$G$-couplings };
\draw[blue,fill=blue!15,thick] (6,0.1)  ellipse (1.4cm and 0.7cm);
\node at (6,0.3) {$H$-couplings};
\draw[red] (7,-0.7)  ellipse (1.4cm and 0.7cm);
\node at (6.5, 1.2)  {$|\mu_{2}-\mu_{3}|=2,4,\ldots$};
\end{tikzpicture}
\caption{Schematic diagram of the relation between $H$- and $G$-couplings }
\label{H and G}
\end{figure}
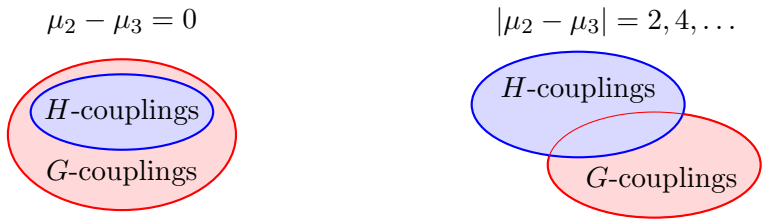

In the case of (A)dS interactions, we have shown that the classifications
of couplings can be entirely rephrased in terms of the possibility of 
rewriting them as $H$-couplings. Hence, the classification 
boils down to the question whether a $G$-coupling (the coupling given by $Y_{i}$'s and $G$) is expressible as a 
$H$-coupling (the coupling given by $H_{ij}$'s and one of $Y_{i}$'s) or not.
Although, our aim was to analyze the three-massless-fields interactions,
let us briefly discuss the interactions of one massless (gauge) field and two massive (matter) fields,
where the distinction between $H$- and $G$-coupling become more clear.
The general solution of the equation \eqref{PDE} consists in two parts:
\be\label{twoclass}
	C(Y,Z)=C_{H}(Y,Z)+\delta_{\mu_{2}-\mu_{3}\in 2\mathbb Z}\ C_{G}(Y,Z)\,.
\ee
The solutions $C_{H}$ correspond to what we call \emph{$H$-couplings}.
Their key property is the existence of a field redefinition 
bringing their form to
\be
	C_{H}(Y,Z)\approx \tilde K(Z_{1},Y_{2},Y_{3},H_{12},H_{13})\,,\label{H sol}
\ee
that it is manifestly off-shell gauge invariant. 
On the other hand, the \emph{$G$-couplings} $C_{G}$ are
present only when $\mu_{2}-\mu_{3}\in 2\,\mathbb Z$\,,
and they may lead to deformations of gauge symmetries.
Their form is given by
\be\label{C G}
	C_{G}(Y,Z)= Y_2^{\sst R\left(\frac{\m_2-\m_3}{2}\right)}\,
	Y_3^{\sst R\left(\frac{\m_3-\m_2}{2}\right)}\,e^{\l
	\,\cD}\,
	K(Z_{1},Y_{1},Y_{2},Y_{3},G)\,\Big|_{G=G(Y,Z)}\,,
\ee
where $R(x)=(x+|x|)/2$ is the ramp function. The latter $H$- and $G$-couplings are not totally independent,
and have a non-vanishing overlap. 
In particular, when \mt{\mu_{2}-\mu_{3}=0}\,,
all the $H$-couplings can be written as $G$-couplings  (see Figure \ref{H and G}).
The physical interpretation of these class of couplings becomes more transparent when one rewrites them as a coupling between a gauge field and a conserved current bilinear in matter fields.
The question whether the deformation is trivial or not becomes the question
whether the current is identically conserved or not.
The $H$-couplings \eqref{H sol} provide us with the full list of interactions 
expressible using identically conserved currents, namely \emph{improvements}. 
On the contrary, 
the couplings in the coset $G/H$ 
correspond to those associated to
on-shell conserved currents, namely \emph{Noether currents}.
It is interesting to notice that Noether currents in (A)dS can involve fields with different masses contrary to the flat-space case.
This means that HS multiplets consisting of matter fields
may involve fields of different masses.
 
As we mentioned, to identify the couplings associated with Noether currents, 
one should quotient the space of $G$-solutions
\eqref{C G} by that of $H$-solutions \eqref{H sol}. 
This can be conveniently done by rewriting them in a common basis
 (see \cite{Joung:2012hz} for more details):
\ba\label{P series basis}
	P^{\t_{1}}_{\s_{1}\s_{2}\s_{3}\u}(\bar\mu_{1};Y,Z) \eq \sum^{p+q\le\s_{1}}_{p,q\ge0}\,\frac{\left[\sigma_2+\frac{\bar\mu_{1}}2\right]_{p}
	\left[\sigma_3-\frac{\bar\mu_{1}}2\right]_q}{\left[\sigma_2+\sigma_3\right]_{p+q}}\,
	\frac{\big(\l\,Z_3\,\partial_{Y_{1}}\,\partial_{Y_2}\big)^{p}}{p!}\,
	\frac{\big(\l\,Z_2\,\partial_{Y_3}\,\partial_{Y_{1}}\big)^{q}}{q!}\times \nn
	&& \hspace{40pt} \times\phantom{\big|}
	Z_{1}^{\,\t_{1}}\,Y_{1}^{\,\s_{1}}\,Y_{2}^{\,\s_{2}}\,Y_{3}^{\,\s_{3}}\,
	[G_{1}(Y,Z)]^{\u}\,,
\ea
where \mt{[a]_{n}:=a(a-1)\cdots(a-n+1)} is the descending Pochhammer symbol and $\bar\m_1=\m_3-\m_2\in 2\,\mathbb Z$\,.
The spins are related to the indices
$\s_{i}, \t_{1}$ and $\u$ by \mt{s_{1}=\s_{1}+\u}\,, 
\mt{s_{2}=\s_{2}+\t_{1}+\u} and
\mt{s_{3}=\s_{3}+\t_{1}+\u}\,.
In this basis,  the genuine Noether couplings
correspond to the ones with \mt{\s_2+\s_3<\s_1}\,. 
Let us also stress that
the allowed mass difference is bounded by spins as
\mt{-2\,s_3\leq\bar \mu_{1}\leq 2\,s_2}.
The corresponding 
gauge-transformation deformation $\delta^{\sst (1)}_{\sst E_{1}}$  of the massive fields are given by the general formula \eqref{T12}. 

\subsection*{Partially-massless fields}

The simple pattern of the non-deforming cubic interactions of massless fields
suggests its straightforward generalization 
to the partially-massless (PM) interactions:
the couplings which do not deform any gauge transformation (the analog of Class I)
have the form:
\be
	C=\sum_{\s_{l-1}=0}^{\m_{l-1}}\sum_{\s_{l+1}=0}^{\m_{l+1}}Y_{l-1}^{\s_{l-1}}Y_{l+1}^{\s_{l+1}}\,\tilde K^{\s_{l-1}\s_{l+1}}(Y_l,\tilde H_1,\tilde H_2,\tilde H_3)\,,\qquad (l=2\,,3)\,,
\ee
while the couplings which deform 
only one of three gauge transformation but no bracket (the analog of Class II)
have the form:
\be
	C=\sum_{\s_{2}=0}^{\m_{2}}\sum_{\s_{3}=0}^{\m_{3}}Y_{2}^{\s_{2}}Y_{3}^{\s_{3}}\,\tilde K^{\s_{2}\s_{3}}(Y_1,\tilde H_1,\tilde H_2,\tilde H_3)\,.
\ee
This conjecture, if true, completes the classification of the PM cubic interactions. 
We do not need to consider a possible distinction like the massless Class III and IV,
since PM representations themselves decomposes into standard massless ones in the flat-space limit. 
In this way, non-Abelian interactions can be identified simply 
by the couplings which cannot be written as above.

\section*{Acknowledgements}

We thank N.~Boulanger, M.~Henneaux, K.~Mkrtchyan and E.~Skvortsov
for useful discussions.
We performed various computations with
 \emph{xAct} \cite{MartinGarcia:2002xa} and \emph{xTras} \cite{Nutma:2013zea} pagkages for Mathematica, 
 and we are grateful to T.~Nutma for his advices. 
 We also acknowledge the GGI, Florence workshop on ``Higher spins, Strings and Duality'' where the present work was developed.

\appendix

{
\section{Transverse and Traceless part}
\label{sec: TT}

In the settings of our works \cite{Joung:2011ww,Joung:2012rv,Joung:2012hz},
one of the most important ingredients
is to consider the TT part of the vertices separately from the other parts.
This enabled us to simplify the technically involved problem of constructing consistent HS interactions into small steps,
of which the first one --- constructing the TT part of the vertices --- could be solved independently from the others.
In this Appendix, we explain this point by recalling the demonstration
provided in \cite{Joung:2012fv}.

In the analysis of the gauge invariance condition $\delta S=0$\,,
it is convenient to split the interacting action $S$ into two parts:
the one which does not involve any divergence, trace or auxiliary fields
(TT part),
and the one which does (DTA part).
The point is that the TT part of the vertex
can be determined  without
using any information about the other part.
First, let us note that
any functional $S$ can be 
separated in a unique way into its TT part and the DTA part as
\be
S=[S]_{\rm\sst TT}+[S]_{\rm\sst DTA}\,,
\ee
after removing all the ambiguities given by integrations by parts or field redefinitions.
Moreover, the gauge invariance condition  can be also split into two equations:
\be\label{N2--}
	\left[\delta^{\sst (0)}S^{\sst (3)}\right]_{\rm\sst TT} \approx 0\,,
	\qquad \left[\delta^{\sst (0)}S^{\sst (3)}\right]_{\rm\sst DTA} \approx 0\,.
\ee
Second, as the gauge variations of 
divergences, traces or auxiliary fields are proportional to themselves
up to terms proportional to $\square$ as
\be
\left[\delta^{\sst (0)}[S^{\sst (3)}]_{\rm\sst DTA}\,
\right]_{\rm\sst TT}\approx 0\,,
\ee
the first equation in \eqref{N2--} provides
an independent condition for the TT parts, $[S^{\sst (3)}]_{\rm\sst TT}$\,, of the interactions:
\be\label{N2---}
	\left[\delta^{\sst (0)}S^{\sst (3)}\right]_{\rm\sst TT}
	=\left[\delta^{\sst (0)}\big\{
	\left[S^{\sst (3)}\right]_{\rm\sst TT}
	+\left[S^{\sst (3)}\right]_{\rm\sst DTA}\big\}\right]_{\rm\sst TT}
	\approx
	\left[\delta^{\sst (0)}\left[S^{\sst (3)}\right]_{\rm\sst TT}\right]_{\rm\sst TT}
	\approx 0\,.
\ee
This analysis is valid for any \emph{homogeneous} condition 
$\delta^{\sst(0)}S^{\sst (n)}\approx 0$
of any order $n$ since 
\mt{\left[\delta^{\sst (0)}[S^{\sst (n)}]_{\rm\sst DTA}\,
\right]_{\rm\sst TT}\approx 0}.
However, it is important to notice that eventual divergence terms 
might contribute to the \emph{inhomogeneous} conditions \eqref{gauge inv}  
when combined with higher-order deformations $\delta^{\sst (n\ge1)}$\,.
Therefore, for higher-order interactions, 
more careful analysis is needed. We hope to report
on this issue in the near future. 

In below, we provide a few more remarks on the issues of 
restricting to the TT part:
\begin{itemize}

\item

Let us recall here an observation made in Section \ref{sec: Global}: 
as far as the global-symmetry algebra 
is concerned, the restriction to the TT part 
does not filter any information since 
gauge parameters associated to global symmetries already satisfy the TT 
conditions.

\item
Let us emphasize once again that in this TT set up,
we are \emph{not} imposing any gauge condition. 
Therefore, the deformation $\delta^{\sst (1)}\Phi$ is 
\emph{not} required to satisfy any condition such as traceless or transverse.
Instead, the consequence of the TT-part restriction 
is that we get only a part --- but all physically relevant --- of the information on $\delta^{\sst (1)}\Phi$\,,
as discussed in the paragraphs after eq.~\eqref{delta 1}.

\item
Finally, let us mention that
there is no vertex whose TT part 
does not induce any deformation while 
its other part does.
It is because our analysis is based on whether a given vertex
can be re-expressed into a $H$ coupling,
whose gauge invariance properties do not rely
on any TT conditions.

\end{itemize}}

\section{Constants generated by $\l$}\label{App: A}

In this appendix, we explain how to eliminate $\l$ in terms of proper factors of 
$\L$\,.
The $\l$\,, which was denoted by $-\hat\delta/L$ in our previous papers\,,
is defined to play the following role inside of the (A)dS action:
\be
	\int_{\rm\sst (A)dS} \l^{n}\,I_{\Delta}
	=\int d^{d+1}X\,\delta^{[n]}\big(\sqrt{\e X^{2}}-L\big)\,\left(\frac{-\e}L\right)^n
	\,
	I_{\Delta}\,,
\ee
where $I_{\Delta}$ is an integrand of the homogeneity degree $\D$\,.
The sign $\e$ is  positive for dS and negative for AdS
and $\delta^{[n]}$ is the $n$-th derivative of the delta distribution.
Hence, after integrating by part the radial variable, one obtains
\be
	\int_{\rm\sst (A)dS} \l^{n}\,I_{\Delta}
	=(\D+d-1)(\D+d-3)\cdots (\D+d-2n+1)\,\L^{n}\,\int_{\rm\sst (A)dS}\,I_{\Delta}\,,
\ee
with the cosmological constant $\L=\epsilon/L^{2}$\,.
Therefore, one can see that $\l^{n}$ inside of action
generates $\L^{n}$ with the above constant
which depends on the space-time dimensionality $d$ and the degree
of homogeneity of the integrand $I_{\D}$\,.
Although the above replacement is possible only inside of the action,
we can still formally assume the above formulas, since the degree of homogeneity $\D$
is preserved through our analysis.

\section{Analysis of trivial deformations}
\label{Appendix: B}

Let us first consider the following simple problem:
for a given function $C$ whether
there exists a solution $\Omega(t)$
for 
\be\label{k Omega}
	(t-Y_{1}\,\partial_{Z_{2}}+Y_{2}\,\partial_{Z_{1}})\,
	\O(t;Y_{1},Y_{2},Z_{1},Z_{2})=C(Y_{1},Y_{2},Z_{1},Z_{2})\,.
\ee
To analyze this, we again expand $\O(t)$ in $t$ as
\be
	\O(t;Y_{1},Y_{2},Z_{1},Z_{2})=
	\sum_{k=0}^{\infty}t^{k}\, \o_{k}(Y_{1},Y_{2},Z_{1},Z_{2})\,,
\ee
then the equation \eqref{k Omega} reduces to
\be
	\o_{-1}=C\,,\qquad
 	\o_{k-1}
	=\left(Y_{1}\,\partial_{Z_{2}}-Y_{2}\,\partial_{Z_{1}}\right) 
	\o_{k}\qquad [\,k=0,1,2,\ldots\,]\,. 
\ee
Since the differential operator commutes with the number operators:
\be
	N_{1}=Y_{1}\,\partial_{Y_{1}}+Z_{2}\,\partial_{Z_{2}}\,,\qquad
	N_{2}=Y_{2}\,\partial_{Y_{2}}+Z_{1}\,\partial_{Z_{1}}\,,
\ee
while have eigenvalues $-1$ with 
\be
	N_{Z}=Z_{1}\,\partial_{Z_{1}}+Z_{2}\,\partial_{Z_{2}}\,,
\ee
we can address this problem in the subspace:
\be
	C\in V^{(n_{1},n_{2},n_{z}-1)}
\ee
where 
\be
	(N_{1},N_{2},N_{Z})\,V^{(n_{1},n_{2},n_{z})}
	=(n_{1},n_{2},n_{z})\,V^{(n_{1},n_{2},n_{z})}\,,
\ee
then the solution belongs to
\be
	\o_{k}\in V^{(n_{1},n_{2},n_{z}+k)}\,.
\ee
The space $V^{\sst (n_{1},n_{2},n_{z})}$ is spanned by
the monomials:
\be
	p^{\sst (n_{1},n_{2},n_{z})}_{m}=
	Y_{1}^{n_{1}-n_{z}+m}\,Y_{2}^{n_{2}-m}\,
	\frac{Z_{1}^{\,m}}{m!}\,\frac{Z_{2}^{\,n_{z}-m}}{(n_{z}-m)!}\,.
\ee
The image of  $\Omega^{\sst (n_{1},n_{2},n_{z})}$ 
by the differential map
is spanned by 
\be
	\Big\{
	p^{\sst (n_{1},n_{2},n_{z}-1)}_{m}-p^{\sst (n_{1},n_{2},n_{z}-1)}_{m-1}\,
	\Big|\,
	{\rm max}\{ n_{z}-n_{1},0\}\le m \le {\rm min}\{n_{z},n_{2}\}\,
	\Big\}
\ee
while the codomain, the space of $C^{\sst (n_{1},n_{2},n_{z}-1)}$\,, 
is spanned by
\be
	\Big\{
	p^{\sst (n_{1},n_{2},n_{z}-1)}_{m}\,\Big|\,
	{\rm max}\{ n_{z}-n_{1}-1,0\}\le m \le {\rm min}\{n_{z}-1,n_{2}\}\,
	\Big\}\,.
\ee
The two sets span the same space if and only if
\be
	n_{z}\le n_{1}  \quad {\rm or} \quad n_{z} \le n_{2}\,.
\ee
In general we deal with funcitons
\be
	C=
	\bigoplus\,C^{\sst (n_{1}-n_{z}+k,n_{2}-n_{z}+k,k-1)}
\ee
satisfy
\be
	\t_{1}+1\le \s_{1} \quad {\rm or} \quad 
	\t_{2}+1\le\s_{2}\,,
\ee	
then there always exists a solution $\Omega$\,.	

\section{Counting of (A)dS structure constants}
\label{Appendix: C}
In this appendix, we give more details on the counting of the number of independent (A)dS structure constants.
As explained in the text, it is not difficult to see, using $\mathfrak{sp}_{2}$ commutation relations, that the general solution for the bracket is given by
\be\label{sp2 ansatz}
	\mathcal F^{s_{3}}{}_{s_{1}s_{2}}=
	e^{Z_1+Z_2}\,\mathfrak F^{s_{3}}{}_{s_{1}s_{2}}(G_{3},H_{3})\,.
\ee
In particular the generic structure constant $\mathcal F^{s_{3}}{}_{s_{1}s_{2}}$ will be a linear combination of the following independent structures:
\be\label{indep terms}
Z_1^{r_2-v-h}\,Z_2^{r_1-v-h}\,G_3^{(v)r_1r_2}\,H_3^h\,,\qquad r_1+r_2-v-2h=r_3\,,
\ee	
where we have used the short-hand notation:
\be
G_3^{(v)r_1r_2}=
\sum_{k=0}^v\,\binom{v}{k}\frac{(Y_1Z_1)^k\,(Y_2Z_2)^{v-k}}{(r_2-v-h+k)!\,(r_1-h-k)!}\, \,.
\ee
Hence, in order to count the independent terms \eqref{indep terms} we need to find the number of solutions of
\be
v+2h=s_1+s_2-s_3-1\,,\qquad v+h\leq \text{min}[s_1,s_2]-1\,,
\ee
that is precisely the condition solved in the text.

\section{No-go for the spin-3 gravitational coupling in flat space}
\label{Appendix: D}

In this appendix, we come back to the discussion on the flat-space gravitational coupling of HS fields of Section~\ref{sec: Global}. The only case left over is
the possibility of allowing the bracket: 
\be
	[\![ \,T_{3}\,,\,T_{2}\,]\!] \sim T_{2}\,,
\ee
induced by the three-derivative $2\!-\!2\!-\!3$ coupling.\footnote{Notice that all $2\!-\!2\!-\!s$ couplings are Abelian for $s>3$ so that the only case in which they might play any role for HS algebras is the $s=3$ case.}
If such a bracket exists, then the RHS of \eqref{Jaco ss2} with \mt{s=s'=3} is not zero
but $\sim T_{2}$ so that one cannot simply conclude $[\![\,T_2\,,\,T_2\,]\!]=0$\,. 
Before analyzing this case in more detail, let us notice that 
this case still excludes any gravitational interactions for spin $s\ge4$ fields,
and requires that gravitons be colored for the non-triviality of $2\!-\!2\!-\!3$ coupling
--- a very exotic scenario, which we shall rule out by the following analysis.
Consider the spin-2 part of the Jacobi identity:
\be
	\big[\hspace{-3pt}\big[\,\big[\hspace{-3pt}\big[
	\,\bar{E}^{\sst (3)}_{1} \,,\,\bar{E}^{\sst (3)}_{2}\,
	\big]\hspace{-3pt}\big]\,,\,\bar{E}^{\sst (2)}_{3}\,\big]\hspace{-3pt}\big]
	+\big[\hspace{-3pt}\big[\,\big[\hspace{-3pt}\big[
	\,\bar{E}^{\sst (3)}_{2} \,,\,\bar{E}^{\sst (2)}_{3}\,
	\big]\hspace{-3pt}\big]\,,\,\bar{E}^{\sst (3)}_{1}\,\big]\hspace{-3pt}\big]
	+\big[\hspace{-3pt}\big[\,\big[\hspace{-3pt}\big[\,
	\bar{E}^{\sst (2)}_{3} \,,\,\bar{E}^{\sst (3)}_{1}\,
	\big]\hspace{-3pt}\big]\,,\,\bar{E}^{\sst (3)}_{2}\,\big]\hspace{-3pt}\big]=0\,,
	\label{jab 332}
\ee
where the superscript $(n)$ indicates the spin of the associated field. 
While the last two terms get contributions only from the 
bracket induced by the \emph{three}-derivative \mt{2\!-\!2\!-\!3} coupling
(hence in total \emph{six} derivatives),
there exist three possible contributions for the first term: 
\begin{enumerate}
\item 
the first bracket by the \emph{three}-derivative \mt{3\!-\!3\!-\!3} coupling; and 
the second bracket by the  \emph{three}-derivative \mt{2\!-\!2\!-\!3} coupling
(hence in total \emph{six} derivatives);
\item 
the first bracket by the \emph{five}-derivative \mt{3\!-\!3\!-\!3} couplings;  and
the second bracket by  the \emph{three}-derivative \mt{2\!-\!2\!-\!3} coupling
(hence in total \emph{eight} derivatives);
\item
the first bracket by the \emph{four}-derivative \mt{2\!-\!3\!-\!3} coupling; and
the second bracket by the \emph{two}-derivative  \mt{2\!-\!2\!-\!2} coupling
(hence in total \emph{six} derivatives).
\end{enumerate}
In fact, the first contribution cannot be considered 
since the three-derivative \mt{3\!-\!3\!-\!3} coupling,
discovered longtime ago \cite{Berends:1984wp},
has been shown to be inconsistent in \cite{Bekaert:2010hp}.
The second contribution is neither possible since 
it involves eight derivatives differently from all other contributions. 
The only remaining possibility is the third case,
and to examine it we need the explicit forms of the brackets:
\be	
	\big[\hspace{-3pt}\big[\,\bar{E}_{1} \,,\,\bar{E}_{2}\,
	\big]\hspace{-3pt}\big]=\cX_{\a}\left(c^{\a}{}_{ab}\,G_3\, H_3\,
	\bar{E}^{\,a}_{1}\,\bar{E}^{\,b}_{2}
	+f^{\a}{}_{\b a}\, H_3\,\bar{E}^{\,\b}_{1}\,\bar{E}^{\,a}_{2}
	+ d^{\a}{}_{\b\g}\,G_3\,\bar{E}^{\,\b}_{1}\,\bar{E}^{\,\g}_{2}\right),
	\label{233 br}
\ee
where the Killing tensors $\bar E_{i}$'s contain both
the spin-2  and spin-3 parts
with Chan-Paton factors labelled respectively
by $\a,\b$ and $a,b$:
\be
	\bar E_{i}=\cX_{\a}\,\bar E^{\,\a}_{i}+\cY_{a}\,\bar E^{\,a}_{i}\,.
\ee
The coefficients $c^{\a}{}_{ab}$, $f^{\a}{}_{\b a}$ and $d^{\a}{}_{\b\g}$ are 
the corresponding structure constants of the internal symmetry.
Plugging eq.~\eqref{233 br} into the Jacobi identity \eqref{jab 332},
one gets 
\ba
&&\tfrac14\,(f^{\a}{}_{\g a}\,f^{\g}{}_{\b b}+
f^{\a}{}_{\g b}\,f^{\g}{}_{\b a})\,Z_{14}
(Y_{12}\, Y_{21}-W_{12}\,Z_{12}
+Y_{13}\, Y_{21}-W_{13}\, Z_{12})
(Y_{23}\, Y_{32} -W_{23}\,Z_{23})\nn
&&+\,d^{\a}{}_{\b \g}\,c^{\g}{}_{ab}\,(
Y_{12}\,Y_{21}-W_{12}\,Z_{12})
(Y_{12}\,Y_{32}\,Z_{24}
+Y_{12}\,Y_{31}\,Z_{24}
+Y_{13}\,Y_{21}\,Z_{34}\nn&&
\hspace{155pt}
-\,Y_{12}\,Y_{23}\,Z_{34}
-Y_{21}\,Y_{31}\,Z_{14}
-Y_{21}\,Y_{32}\,Z_{14})=0\,,
\label{sim jab}
\ea
where $(W_{ij},Y_{ij},Z_{ij})=
(\partial_{X_{i}}\!\cdot\partial_{X_{j}},
\partial_{U_{i}}\!\cdot\partial_{X_{j}},
\partial_{U_{i}}\!\cdot\partial_{U_{j}})$\,,
and the equality holds when it acts on
the Killing tensors
$\bar{E}^{\,a}_{1}\,\bar{E}^{\,b}_{2}\,\bar{E}^{\,\b}_{3}$\,.
Apparently, this expression as a function of the variables 
$W_{ij},Y_{ij},Z_{ij}$'s is not identically zero.
However, one should examine all possible identities of 
such variables due to the properties of Killing tensors.
For that, we have evaluated the expression on the generators of the type:
\be
\bar{E}^{\sst (n)}_{i}=\left[(W_{i})_{MN}\,U^M\,X^N\right]^n\,,
\ee
with antisymmetric constant matrices $W_{i}$'s.
This has been carried out with the help of Mathematica  
and we found that the relation \eqref{sim jab}
cannot hold for non-trivial internal 
structure constants $c^{\a}{}_{ab}$, $f^{\a}{}_{\b a}$ and $d^{\a}{}_{\b\g}$\,.

\bibliographystyle{JHEP}
\bibliography{ref}

\end{document}